\RequirePackage{fix-cm}
\documentclass[smallcondensed]{svjour3}       
\smartqed  
\usepackage{appendix}
\usepackage{amsmath}
\usepackage{graphicx}
\usepackage{lineno}
\usepackage{array}
\usepackage{longtable}
\usepackage{natbib}
\setcitestyle{aysep={}}

\newcommand*\patchAmsMathEnvironmentForLineno[1]{%
\expandafter\let\csname old#1\expandafter\endcsname\csname #1\endcsname
\expandafter\let\csname oldend#1\expandafter\endcsname\csname end#1\endcsname
\renewenvironment{#1}%
{\linenomath\csname old#1\endcsname}%
{\csname oldend#1\endcsname\endlinenomath}}%
\newcommand*\patchBothAmsMathEnvironmentsForLineno[1]{%
\patchAmsMathEnvironmentForLineno{#1}%
\patchAmsMathEnvironmentForLineno{#1*}}%
\AtBeginDocument{%
\patchBothAmsMathEnvironmentsForLineno{equation}%
\patchBothAmsMathEnvironmentsForLineno{align}%
\patchBothAmsMathEnvironmentsForLineno{flalign}%
\patchBothAmsMathEnvironmentsForLineno{alignat}%
\patchBothAmsMathEnvironmentsForLineno{gather}%
\patchBothAmsMathEnvironmentsForLineno{multline}%
}

\begin{document}

\title{Comparing phase-space and phenomenological modeling approaches for Lagrangian particles settling in a turbulent boundary layer}

\author{Andrew P. Grace         \and
        David H. Richter \and Andrew D. Bragg
}

\institute{Andrew P. Grace \at
            Civil and Environment Engineering and Earth Sciences, \\
            University of Notre Dame,\\
            Notre Dame, Indiana 46556, USA\\
            \email{agrace4@nd.edu}           
            \and
            David H. Richter \at
            Civil and Environment Engineering and Earth Sciences, \\
            University of Notre Dame,\\
            Notre Dame, Indiana 46556, USA
            \and
            Andrew D. Bragg \at
            Department of Civil and Environmental Engineering, \\ 
            Duke University, \\
            Durham, North Carolina 27708, USA
}

\date{Received: DD Month YEAR / Accepted: DD Month YEAR}

\maketitle

\begin{abstract}
It has long been known that under the right circumstances, inertial particles (such as sand, dust, pollen, or water droplets) settling through the atmospheric boundary layer can experience a net enhancement in their average settling velocity due to their inertia. Since this enhancement arises due to their interactions with the surrounding turbulence it must be modelled at coarse scales. Models for the enhanced settling velocity (or deposition) of the dispersed phase that find practical use in mesoscale weather models are often \textit{ad hoc} or are built on phenomenological closure assumptions, meaning that the general deposition rate of particle is a key uncertainty in these models. Instead of taking a phenomenological approach, exact phase space methods can be used to model the physical mechanisms responsible for the enhanced settling, and these individual mechanisms can be estimated or modelled to build a more general parameterization of the enhanced settling of inertial particles. In this work, we use direct numerical simulations (DNS) and phase space methods as tools to evaluate the efficacy of phenomenological modelling approaches for the enhanced settling velocity of inertial particles for particles with varying friction Stokes numbers and settling velocity parameters. We use the DNS data to estimate profiles of a drift-diffusion based parameterization of the fluid velocity sampled by the particles, which is key for determining the settling velocity behaviour of particles with low to moderate Stokes number. We find that by increasing the settling velocity parameter at moderate friction Stokes number, the magnitude of preferential sweeping is modified, and this behaviour is explained by the drift component of the aforementioned parameterization. We then use these profiles to argue that the eddy-diffusivity-like closure used in phenomenological models is incomplete, relying on inadequate empirical corrections. Finally, we discuss opportunities for reconciling exact phase space approaches with simpler phenomenological approaches for use in coarse-scale weather models.
\end{abstract}

\section{Introduction}

A detailed and fundamental understanding of the processes responsible for the dispersion of dust and sand particles (about 0.5 - 100 $\mu$m in size \citep{kok_physics_2012,shao_dust_2008}) in the atmospheric boundary layer is of utmost importance for accurate modeling of their impact on biogeochemical cycles \citep{ryder_coarse-mode_2018}, global radiation balance \citep{miller_mineral_2006}, air quality \citep{querol_monitoring_2019}, and a host of other processes. More fundamentally, the interactions between dust particles 
and turbulent air presents an interesting and important multi-scale physics and engineering problem. Dust and sand particles are emitted from the ground though a series of complex interactions between gravity and the wind stresses from synoptic-scale events, electrostatic forces, and dynamic forces \citep{kok_physics_2012}, and under certain circumstances, may rise high into the atmosphere \citep{kok_smaller_2017} and be transported thousands of kilometers away \citep{shao_dust_2008} (depending on their size). Part of the challenge of predicting the global transport in the atmosphere is accurate prediction of the rate at which particles of different sizes settle out of the atmosphere under the action of gravity (through their Stokes settling velocity), and how different scales of turbulent motion in the planetary boundary layer (PBL) either enhance or suppress this process. 
 

It has long been known that under net settling conditions, heavy particles settling through turbulence can experience a net enhancement in their average settling velocity, and many studies have linked this to their inertia \citep{wang_settling_1993,aliseda_effect_2002,good_settling_2014,brandt_particle-laden_2022}. A particle's inertia is quantified by its Stokes number, which is the ratio between the particle's relaxation timescale, and a fluid timescale. A related parameter is the settling velocity parameter, which is the ratio of the Stokes settling velocity to a characteristic fluid velocity. Settling velocity enhancement comes from a variety of sources, but pertinent to the work presented here is the preferential sweeping mechanism \citep{wang_settling_1993,tom_multiscale_2019}, which says that when the turbulence is locally homogeneous and there are no mean gradients in the turbulence properties, inertial particles settling under the force of gravity preferentially sample the downward side of turbulent eddies. This means that on average, the fluid velocities they sample are negative, which leads to an enhancement of the average settling velocity above what would be predicted by the Stokes settling velocity. It can be difficult to observe the enhanced settling velocity due to preferential sweeping in the natural environment, but nevertheless, \citet{nemes_snowflakes_2017} and \citet{li_settling_2021} presented field observations of snowflakes falling through a neutral boundary layer that had fall speeds much higher than their estimated Stokes settling velocity, and they argued that the enhancement was due to preferential sweeping. 

Understanding and being able to predict the magnitude of this effect as a function of Stokes number and settling velocity parameter is of utmost importance for accurate modeling of both the horizontal and vertical dispersion of particles and aerosols. 
Since the magnitude of the interactions between the particles and turbulent eddies are not known \textit{a priori}, they must be modelled. On the one hand, in controlled numerical and laboratory experiments, empirical expressions for the enhanced settling velocity in terms of the Stokes number, settling parameter and Reynolds number have been proposed for homogeneous isotropic turbulence \citep{rosa_settling_2016}. Meanwhile, models for the enhanced settling velocity (or deposition) of the dispersed phase that find practical use in mesoscale weather models (a list of which can be found in \citet{kukkonen_review_2012}) are often quite simple and \textit{ad hoc} or are built on phenomenological closure assumptions (such as an eddy diffusivity approximation), meaning that the settling rate of particle is a key uncertainty in these models in general. For example, models in use often take a crude approach, which, roughly speaking, represent the enhanced settling velocity as a sum of the Stokes settling velocity and an additive correction based on the collection efficiencies of various land use categories \citep{slinn_predictions_1980,slinn_predictions_1982,zhang_size-segregated_2001,emerson_revisiting_2020,farmer_dry_2021}. These approaches are known as resistance-based approaches, and are discussed at length in \citet{seinfeld_atmospheric_1998}. 

Instead of these \textit{ad hoc} resistance approaches, others have developed relationships between the surface flux and mean concentration profile which are based on the conservation equations for particle concentration. In the limit of assuming that the diffusive effects of turbulence on the particles is in exact balance with gravitational settling, one arrives at the so-called ``Rouse profile'' \citep{rouse_modern_1937,prandtl_essentials_1952,boudreau_rouse_2020}. This simple power-law profile has been extended for non-equilibrium conditions (meaning a net constant downward or upward flux)\citep{hoppel_particle_2005,kind_one-dimensional_1992} and non-neutral atmospheric stability\citep{freire_flux-profile_2016}. The Rouse profile serves as a yardstick by which to compare high Reynolds number laboratory experiments \citep{berk_transport_2020} and direct numerical simulations \citep{richter_inertial_2018}, but is only accurate in the limit of vanishing inertia. 
Corrections to account for the effect of particle inertia in the near-wall region (the viscous sub-layer) have been applied to the mass conserving approach. The common application is to practically model a process known as impaction, which is the term used to refer to the rate at which settling particles impact a canopy element based on their ability to adjust to changes in fluid streamlines, which is a function of the particle inertia \citep{emerson_revisiting_2020}.
Thus the degree to which inertia is relevant to a particle's settling behaviour in phenomenological models is often contained within the impaction component of the model. In order to parameterize the effect of different land use categories on the settling behaviour (such as grasslands, forests, and water surfaces, for example) empirical corrections and significant model calibration are required, as the details can vary significantly depending on the land use category \citep{zhang_size-segregated_2001,farmer_dry_2021}.

As an alternative to the phenomenological approaches discussed above, exact phase space methods which quantify the evolution of the probability density function of an ensemble of particles, such as that outlined in \citep{pope_turbulent_2000}, can be used to explicitly represent the settling velocity as the sum of distinct mechanisms which can be estimated or modelled to build a more general parameterization of the enhanced settling of inertial particles. In a recent paper, \citet{bragg_mechanisms_2021} used this approach to identify the important mechanisms governing particle settling in coupled direct numerical simulations of a turbulent boundary layer and Lagrangian point particles for negative net flux (settling) conditions. They computed averaged profiles of each of the terms, confirming that for settling conditions within the interior of the turbulent boundary layer (what they refer to as the quasi-homogeneous region), the dominant settling mechanisms for low and moderate Stokes number were the Stokes settling velocity and preferential sweeping. At higher Stokes number, they additionally identified turbophoresis as an important mechanism. Finally, they discussed the necessity of a drift-diffusion type closure \citep{reeks_continuum_1992,skartlien_kinetic_2007,bragg_particle_2012} for modeling the fluid velocities sampled by the settling particles, though they do not quote an exact form.

In this paper, we take inspiration from \citet{bragg_mechanisms_2021} and discuss and compare two different approaches for modeling the settling of inertial particles through turbulent boundary layers at different friction Stokes numbers ($\mathrm{St}^+$) and settling velocity parameters ($\mathrm{Sv}^+$). The first approach is based on the phase space probability of the ensemble of particles discussed in both \citet{bragg_mechanisms_2021} and \citet{skartlien_kinetic_2007}. This method is exact but unclosed; we adopt the drift-diffusion closure discussed in \citet{reeks_continuum_1992} to model the fluid velocities sampled by the particle. We then use the DNS data to assist in estimating the magnitude and the overall behaviour of the components of this closure across the quasi-homogeneous region of the turbulent boundary layer for varying $\mathrm{St}^+$ and $\mathrm{Sv}^+$. We then compare the phase-space approach to a phenomenological but mass-conserving approach discussed in \citet{hoppel_particle_2005} and \citet{giardina_new_2018}. This approach is similar to the \textit{ad hoc} resistance based approaches discussed earlier. The phenomenological approach is based on an eddy-diffusivity closure for the turbulent fluxes of particles, which allows us to solve the governing differential equation for an equation for average vertical particle velocity at a reference height. We show that while these \textit{ad hoc} approaches might provide reasonable deposition estimates, the approach they take is fundamentally inappropriate as they are assuming \textit{a priori} that the turbulent fluxes can be modelled specifically as a diffusive process, and not one that includes a drift component. We then discuss the implications of the different approaches for modeling the settling of inertial particles and recommend improvements that have potential to be implemented in operational models.

In section \ref{setup} we introduce the numerical model, and the equations of motion for both the carrier phase and the dispersed phase, and finally outline the series of Numerical Experiments for the rest of the work. Section \ref{models} outlines the phase-space and closure assumption that will be employed in the paper, as well as details regarding the phenomenological modeling approach that will serve as a comparative lens by which to view the results. Section \ref{results} highlights the results for the two Numerical Experiments described in section \ref{setup}, and finally section \ref{discussion} provides a brief summary of the results and a discussion of the comparison between the two modeling approaches.  

\section{Model setup \label{setup}}
\subsection{Carrier Phase}
\begin{figure}
    \centering
    \includegraphics[width=\textwidth]{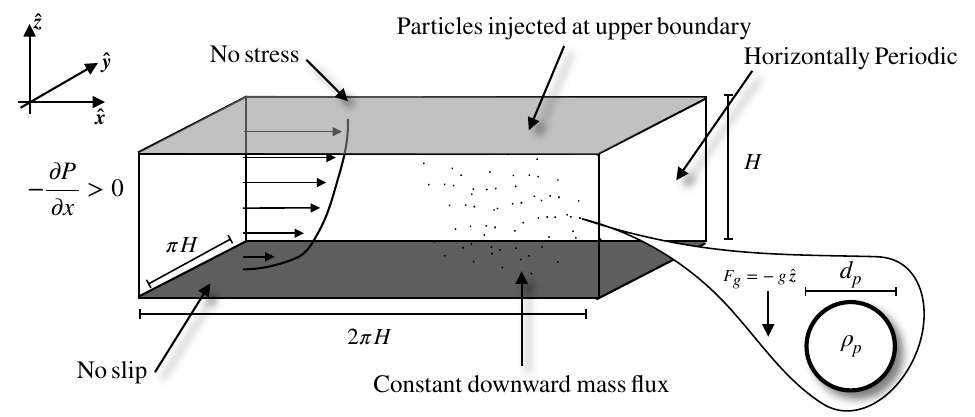}
    \caption{The domain is a rectangular channel of height $H$, streamwise length $2\pi H$ and spanwise width $\pi H$. The carrier phase is subjected to a constant pressure gradient in the streamwise direction, and the boundary conditions enforced at the top and bottom boundaries are no-stress and no-slip respectively, while the horizontal boundary conditions are periodic. Particles are injected at the upper boundary at a random horizontal location with an initial velocity equal to $\tau_pg$ and removed when they contact the bottom boundary. They are allowed to rebound elastically off the upper boundary.}
    \label{fig:schematic}
\end{figure}

In this work, we use the NCAR Turbulence with Lagrangian Particles Model (NTLP) to model one-way coupled inertial particles settling through a turbulent boundary layer, which has been used for numerous particle-laden turbulence studies \citep{richter_inertial_2018,wang_inertial_2019,gao_direct_2023}. For the carrier phase, we use direct numerical simulations (DNS) to solve the three dimensional incompressible constant density Navier-Stokes equations in a turbulent open channel flow setup:
\begin{align}
    \frac{D\boldsymbol{u}}{Dt} &= -\frac{1}{\rho_a}\nabla p + \nu\nabla^2\boldsymbol{u} - \frac{1}{\rho_a}\frac{d P}{dx}\hat{\boldsymbol{x}}, \\
    \nabla \cdot \boldsymbol{u} &= 0.
\end{align}
A schematic of the setup is presented in figure \ref{fig:schematic}.
In the above equations, $\frac{D}{Dt}$ represents the material derivative, $\boldsymbol{u}$ represents the three dimensional flow velocity, $p$ represents the turbulent pressure field, $\rho_a$ is the air density, and $\nu$ is the kinematic viscosity. The flow is one-way coupled, meaning that the flow does not feel the effects of the dispersed phase. 

At the lower boundary, a no-slip boundary condition is enforced, while at the upper boundary, a no-stress boundary condition is enforced. The domain is periodic in the $x$ and $y$ directions. The background state of the carrier phase is established by accelerating the flow with an imposed pressure gradient, $-dP/dx>0$ (note that $\hat{\boldsymbol{x}}$ is the unit vector in the streamwise direction) and allowing the flow to become turbulent. The magnitude of the pressure gradient allows us to define a friction velocity $u_\tau = \sqrt{\tau_w/\rho_a}$, where $\tau_w$ is the stress at the lower boundary. Using the friction velocity, the height of the domain, $H$, and viscosity of the carrier phase, we can define a friction Reynolds number of $\mathrm{Re}_\tau = \frac{u_\tau H}{\nu}$. Friction Reynolds numbers for each simulation presented in this work can be found in tables \ref{tab:cases} and \ref{tab:cases_slinn}.

\subsection{Dispersed Phase}
For each particle (the dispersed phase), we apply the point-particle approximation and solve the conservation of momentum for a rigid spherical particle subjected to linear hydrodynamic drag and the gravitational force: 
\begin{eqnarray}
    \frac{d\boldsymbol{v}_p}{dt} =& &\frac{1}{\tau_p}\left(\boldsymbol{u}_p - \boldsymbol{v}_p\right) - \boldsymbol{g}, \label{newton 2} \\ 
    \frac{d\boldsymbol{x}_p}{dt} =& &\boldsymbol{v}_p. \label{part pos}
\end{eqnarray}
Here, $\boldsymbol{v}_p$ represents the three dimensional velocity vector of each particle, $\boldsymbol{x}_p$ is the location of each particle, $\boldsymbol{g}$ is the gravitational acceleration (which only affects accelerations in the $z$ direction), $\boldsymbol{u}_p$ is the flow velocity evaluated at the location of the particle, and $\tau_p$ is the relaxation timescale of the particle, defined as 
\begin{equation}
    \tau_p = \frac{\rho_pd_p^2}{18\rho_0\nu},
\end{equation}
where $\rho_p$ is the particle density and $d_p$ is the particle diameter.

Once the system is at equilibrium, a vertically constant mass flux is maintained by injecting Lagrangian particles at the upper boundary of the domain with a downward vertical velocity equal to their Stokes settling velocity ($\tau_p g$) at a random horizontal location. Particles are removed from the domain once they contact the bottom boundary, ensuring that a new particle is re-injected after one is removed thereby maintaining a constant particle number. Particles are also allowed to rebound elastically from the upper boundary, though this happens rarely. 

Using the friction velocity and the kinematic viscosity of the carrier phase, we can non-dimensionalize the particle diameter, particle relaxation timescale, and the Stokes settling velocity;
\begin{equation}
    d^+ = \frac{d_pu_\tau}{\nu},\quad \mathrm{St}^+ = \frac{\tau_pu_\tau^2}{\nu},\quad \mathrm{Sv}^+ = \frac{\tau_pg}{u_\tau}.
\end{equation}
For the rest of this study, the normalized particle diameter (which can also be interpreted as a particle Reynolds number) is held fixed at $d^+ = 0.236$. $\mathrm{St}^+$ is a viscous Stokes number and quantifies the inertia of the particle. $\mathrm{Sv}^+$ is the settling velocity parameter, which with this definition is directly related to the Rouse number from Rouse-Prandtl theory \citep{rouse_modern_1937,prandtl_essentials_1952}. Finally, we will also refer to the integral Stokes number $\mathrm{St} = \tau_p/\tau$, where $\tau$ is the Lagrangian integral timescale (also known as the Lagrangian timescale of turbulence). For a turbulent boundary layer, the Lagrangian integral timescale increases with height as turbulent motions higher in the boundary layer are highly correlated with themselves for longer. As an example, see the work of \citet{oesterle_lagrangian_2004}. More information on the computation of this parameter can be found in section \ref{phasespace}.

\subsection{Numerical Experimental setup}
In this manuscript, we take two Numerical Experimental approaches. The first is to vary $\mathrm{St}^+$ and $\mathrm{Sv}^+$ independently, and parameter values for these cases can be found in table \ref{tab:cases}. This has been done in several studies focusing on the dominant mechanisms responsible for inertial particle settling in turbulent boundary layers \citep{bragg_settling_2021,bragg_mechanisms_2021} and homogeneous isotropic turbulence \citep{wang_settling_1993,rosa_settling_2016}. Practically speaking, to choose $\mathrm{St}^+$, we vary the particle density, $\rho_p$, whereas to choose $\mathrm{Sv}^+$, we vary the magnitude of the gravitational acceleration, $g$. Thus, in addition to the relaxation timescale $\tau_p$, there is a second timescale associated with $\mathrm{Sv}^+$ that is proportional to $\tau_s = \sqrt{\ell g^{-1}}$, where $\ell$ is some length scale. 

The second experimental approach we take is to couple $\mathrm{St}^+$ and $\mathrm{Sv}^+$ though a parameter we denote as $\Gamma$. When studying particle deposition processes in geophysical contexts (i.e. deposition over the ocean, glaciers, tundras, forest canopies, etc.), $g$ is not a free parameter, and $\mathrm{Sv}^+$ and $\mathrm{St}^+$ are not independent but are instead linked through characteristics of the turbulent flow (i.e. a known $\nu$, $u_\tau$, and $g$). Thus a particle's inertia and its settling velocity are physically coupled, and can be written: $\mathrm{Sv}^+ =\Gamma \mathrm{St}^+$, where $\Gamma = \nu g/u_\tau^3$. In order to make a comparison between our work and pre-existing work on the deposition of inertial particles, we include several cases where $\mathrm{St}^+$ and $\mathrm{Sv}^+$ are not varied independently.
Parameter values for these cases are shown in table \ref{tab:cases_slinn}.

The cases in table \ref{tab:cases} were chosen to have $\mathrm{Re}_\tau = 630$, while the cases in table \ref{tab:cases_slinn} have $\mathrm{Re}_\tau = 315$, which is the same $\mathrm{Re}_\tau$ as the simulations presented in \cite{bragg_mechanisms_2021}. It is well known that a there is a significant portion of the flow at both these Reynolds numbers that exhibit the traditional logarithmic behaviour in the quasi-homogeneous layer \citep{kantha_small_2000,bragg_mechanisms_2021}, so we do not expect any significant Reynolds number dependence between the two sets of cases.

\begin{table}
\centering
\begin{tabular}{ccccc}
$\mathrm{Re}_\tau$ & $\mathrm{Sv}^+$  & $\mathrm{St}^+$  & $\Gamma$ & $N_x\times N_y\times N_z$ \\ \hline \hline
630 & 0.025   & 0.1  & 0.25   & $256 \times 256\times 256$    \\ 
630 & 0.025   & 10  & 0.0025   & $256 \times 256\times 256$    \\ 
630 & 0.025   & 100  & 0.00025   & $256 \times 256\times 256$    \\ \hline \hline
630 & 0.1   & 0.1  & 1   & $256 \times 256\times 256$    \\ 
630 & 0.1   & 10  & 0.01   & $256 \times 256\times 256$    \\ 
630 & 0.1   & 100  & 0.001   & $256 \times 256\times 256$    \\ \hline \hline
630 & 0.25   & 0.1  & 2.5  & $256 \times 256\times 256$    \\ 
630 & 0.25   & 10  & 0.025   & $256 \times 256\times 256$    \\ 
630 & 0.25   & 100  & 0.0025   & $256 \times 256\times 256$    \\ \hline \hline
\end{tabular}
\caption{The above table highlights the experiments discussed in section \ref{phasespace}. We will refer to this set of simulations as ``Numerical Experiment 1". $\mathrm{Re}_\tau$ is the friction Reynolds number, $\mathrm{Sv}^+$ is the settling velocity parameter in terms of viscous units, $\mathrm{St}^+$ is the Stokes number in terms of viscous units, $\Gamma$ is the ratio of $\mathrm{Sv}^+$ and $\mathrm{St}^+$, and $N_x\times N_y\times N_z$ are the number of grid points in the streamwise, spanwise, and vertical directions respectively. For this experiment, $\mathrm{St}^+$ and $\mathrm{Sv}^+$ are independently varied. \label{tab:cases}}
\end{table}

\begin{table}[htbp!]
\centering
\begin{tabular}{ccccc}
$\mathrm{Re}_\tau$ & $\mathrm{Sv}^+$  & $\mathrm{St}^+$  & $\Gamma$ & $N_x\times N_y\times N_z$ \\ \hline \hline
315 & 0.00025 & 0.1 & 0.0025   & $128 \times 128\times 128$   \\
315 & 0.0025 & 1   & 0.0025   & $128 \times 128\times 128$    \\
315 & 0.025 & 10  & 0.0025   & $128 \times 128\times 128$    \\ 
315 & 0.25 & 100  & 0.0025   & $128 \times 128\times 128$    \\ \hline
315 & 0.0001   & 0.1 & 0.001   & $128 \times 128\times 128$    \\
315 & 0.001   & 1   & 0.001   & $128 \times 128\times 128$   \\
315 & 0.01   & 10  & 0.001   & $128 \times 128\times 128$    \\ 
315 & 0.1   & 100  & 0.001   & $128 \times 128\times 128$    \\ \hline \hline
\end{tabular}
\caption{The above table highlights the experiments discussed \ref{resistance}. We will refer to this set of simulations as ``Numerical Experiment 2". The Parameters take the same definitions as those discussed in table \ref{tab:cases}. Note that in these cases $\mathrm{St}^+$ and $\mathrm{Sv}^+$ are coupled, meaning that $\Gamma$ is held constant for each subset of cases. \label{tab:cases_slinn}}
\end{table}

\section{Models of particle deposition \label{models}}
In this manuscript, we will compare two modeling approaches for the deposition of inertial particles in a turbulent boundary layer, with specific interest in the dominant behaviour in the log layer. This layer is often taken to represent to lowest 100 meters of the atmospheric boundary layer, and its characteristics are instrumental in determining the dominant processes which control inertial particle settling. The first approach begins from the phase space equation of the joint probability density function of the particles and will be the main approach we will discuss. This approach has the benefit of introducing no additional approximations other than the ones already found in the equations describing the motion of the dispersed and carrier phases \citep{bragg_mechanisms_2021}. The second model, which will be our point of comparison, is based off of a common eddy diffusivity closure found in the dry deposition literature \citep{hoppel_surface_2002,hoppel_particle_2005,giardina_new_2018,farmer_dry_2021}. This method has found practical and operational use but relies on a phenomenological closure scheme necessarily requiring empirical corrections. These methods take different approaches to estimate the same quantity and the goal of this paper is to compare contexts where they are both successful as well as each of their shortcomings.

In the following sections, we will show both the phase-space model and the phenomenological closure models, omitting some details for brevity and referring to external literature when reasonable. 

\subsection{Phase space approach \label{phasespace}}

We can define the joint PDF in position-velocity phase space, $\mathcal{P}$, as
\begin{equation}
    \mathcal{P} = \langle\delta(\boldsymbol{x}_p - \boldsymbol{x})\delta(\boldsymbol{v}_p - \boldsymbol{v})\rangle,
\end{equation}
where $\boldsymbol{x}_p$ and $\boldsymbol{v}_p$ are the time-dependent particle position and velocity relative to the origin, while $\boldsymbol{x}$ and $\boldsymbol{v}$ are the time-independent position and velocity coordinates of the phase space in which the particle motion is being described. Here, $\delta(\cdot)$ is the Dirac delta function and $\langle\cdot\rangle$ denotes an ensemble average over all realizations of the system. Taking a time derivative of $\mathcal{P}$, we can form an evolution equation of the phase space probability density for the particles. 

\begin{equation}
    \frac{\partial \mathcal{P}}{\partial t} + \nabla_{\boldsymbol{x}}\cdot(\langle\boldsymbol{v}\rangle_{\boldsymbol{x},\boldsymbol{v}}\mathcal{P})
 + \nabla_{\boldsymbol{v}}\cdot(\langle\dot{\boldsymbol{v}}\rangle_{\boldsymbol{x},\boldsymbol{v}}\mathcal{P}) = 0,
\end{equation}
where $\dot{\boldsymbol{v}}_p$ is the time derivative of the velocity, and $\langle\cdot\rangle_{\boldsymbol{x},\boldsymbol{v}}$ denotes an ensmble average conditioned on $\boldsymbol{x}_p(t)=\boldsymbol{x},\boldsymbol{v}_p(t)=\boldsymbol{v}$. The operators $\nabla_{\boldsymbol{x}}$ and $\nabla_{\boldsymbol{v}}$ represent derivatives with respect position and velocity respectively. To obtain a governing equation for the number concentration of the particles, we can integrate the above equation over all velocities:

\begin{equation}
    \frac{\partial \varrho}{\partial t} + \nabla_{\boldsymbol{x}} \cdot \langle{\boldsymbol{v}_p}\rangle_{\boldsymbol{x}}\varrho = 0, \label{unsteady number concentration evolution}
\end{equation}
where the particle number concentration is defined by
\begin{equation}
    \varrho(\boldsymbol{x},t) = \int \mathcal{P}d\boldsymbol{v}, \label{number concentration}
\end{equation}
and the particle mean momentum is defined by
\begin{equation}
    \langle{\boldsymbol{v}_p}\rangle_{\boldsymbol{x}}\varrho = \int \boldsymbol{v}\mathcal{P}d\boldsymbol{v}. \label{particle momentum}
\end{equation}
Equations \eqref{number concentration} and \eqref{particle momentum} are consequences of the sifting property of the Dirac delta function.
Thus, at steady state, the governing equation for the number concentration is:
\begin{equation}
    \nabla_{\boldsymbol{x}} \cdot \langle{\boldsymbol{v}_p}\rangle_{\boldsymbol{x}}\varrho  = 0.
\end{equation}
By assuming a horizontally periodic domain, the above equation simplifies to 
\begin{equation}
    \frac{\partial}{\partial z}\left(\varrho \langle w_p \rangle_z\right) = 0,
\end{equation}
where $\langle w_p \rangle_z$ is the average vertical component of the particle velocity conditioned on $z_p(t)=z$.
Thus, we arrive at 
\begin{equation}
    \varrho \langle w_p \rangle_z = F, \label{phase space settling}
\end{equation}
where $F$ is the constant vertical number flux of the inertial particles, thereby giving us an algebraic relationship between the averaged vertical particle velocity, the total flux, and the horizontally averaged number concentration profile. Note that in practice, for the statistically stationary, horizontally homogeneous system we are considering, the conditional averages $\langle \cdot \rangle_z$ can be evaluated by spatial averaging over the homogeneous directions, as well as averaging in time.

At this point, we have not yet introduced a model for the average settling velocity, $\langle w_p\rangle_z$. To do so, we can form an evolution equation for the first moment of the joint PDF given by \eqref{particle momentum}. This is done rigorously in previous work for particle momentum equations that contain hydrodynamic drag and gravity (see \citet{skartlien_kinetic_2007,bragg_mechanisms_2021}), so the exact algebraic details will be omitted here. After deriving an equation for the first moment we arrive at the following equation for a statistically stationary, horizontally homogeneous flow
\begin{equation}
    \varrho\langle w_p\rangle_z = \varrho\langle u_p\rangle_z -\varrho \tau_p g - \tau_p\varrho\frac{\partial}{\partial z}\left(\frac{1}{2}\langle w_p\rangle_z^2\right) -\tau_p\frac{\partial}{\partial z}\varrho S, \label{settling evolution}
\end{equation}
where $\langle u_p\rangle_z$ represents the average vertical fluid velocities sampled by particles at height $z_p(t)=z$, and $S$ represents the variance of the vertical particle velocity for particles at height $z_p(t)=z$. 

The above equation decomposes the mean particle momentum, $\varrho\langle w_p\rangle_z$, into contributions from various mechanisms, and the mechanisms contained in \eqref{settling evolution} are a consequence of the particle equations of motion, \eqref{newton 2} and \eqref{part pos}. The first term on the right hand side \eqref{settling evolution} is the contribution to the vertical flux associated with the average vertical fluid velocity sampled by the particles, $\langle u_p\rangle_z$. The second term on the right hand side is the contribution associated with Stokes settling velocity. The third term on the right hand side arises from the average wall normal convective acceleration of the particles, but was shown by
\cite{bragg_mechanisms_2021} to be negligible in the settling regime this work is concerned with. Finally, the fourth term on the right hand side is actually two separate mechanisms. They are a diffusive flux arising from the partial decoupling of particle and fluid velocities and the turbophoretic drift velocity arising from vertical variation in the turbulence intensity. The magnitudes of these terms in a turbulent boundary layer are discussed at length in \citet{bragg_mechanisms_2021}. Since these two mechanisms are not the main focus of the paper, they are grouped together for notational simplicity.

As mentioned previously, the primary concern of this work is related to the estimate of the average settling velocity in the logarithmic region of a turbulent boundary layer (i.e. $z^+> 100$). In this layer, the gradients of mean properties are relatively weak, so to leading order, we expect that for particles with low to moderate inertia, the dominant contribution to the vertical flux is associated with the fluid velocities sampled by the particles. However, for $\mathrm{St}\rightarrow 0$ (where $\mathrm{St}$ is a Stokes number based on an integral scale of the flow), we expect particles to exactly follow fluid streamlines, so particles become passive tracers. Likewise, in the limit of $\mathrm{St}\rightarrow \infty$, the particle trajectories are no longer correlated with the flow. Thus, in these limits and particles uniformly sample flow velocities.

Since $\langle u_p\rangle_z$ is a flow property, it is not known \textit{a priori}, so it must be estimated. We can use the following closure approximation for the fluid velocity sampled by the particle \citep{reeks_continuum_1992}:
\begin{equation}
    \varrho \langle u_p\rangle_z \approx -\tau_p\left(\gamma + \frac{\partial\lambda}{\partial z}\right) \varrho - \tau_p\lambda \frac{\partial \varrho}{\partial z}, \label{closure}
\end{equation}
where $\gamma$ and $\lambda$ are functions that depend on the two-point, two-time correlations of the fluid velocity along the particle trajectory. As discussed in detail in \citet{bragg_drift-free_2012}, various other closure approximations for $\varrho \langle u_p\rangle_z$ have been derived using different approaches that all have the form of \eqref{closure} but differ in the details regarding how $\gamma$ and $\lambda$ are defined. These differences, while of interest from the perspective of mathematical rigor, are not important for the present study. We choose the form presented in \cite{reeks_continuum_1992} because its specification for $\gamma$ and $\lambda$ can be written in the simple forms:
\begin{eqnarray}
    \tau_p\lambda &= \langle \Delta z u_p(z,t)\rangle, \label{lambda_correlation} \\ 
    \tau_p\gamma &= -\left< \Delta z \frac{\partial u_p}{\partial z}(z,t) \right>, \label{gamma correlation} 
\end{eqnarray}
where the accumulated vertical displacement due to the fluid velocities sampled by the fluid over its path from a time $t_1$ to the measurement time $t$ is written as 
\begin{equation}
    \Delta z = \int_{t_1}^t u_p(z,t;s)\frac{1}{\tau_p}\left(1 - e^{(s - t)/\tau_p}\right)ds,
\end{equation}
which is discussed in detail in \citet{skartlien_kinetic_2007}.
Equation \eqref{lambda_correlation} is interpreted as the correlation between changes in the vertical position of the particles and the velocities they sample along their trajectories, whereas \eqref{gamma correlation} is the correlation between changes in the vertical position of the particles and the fluid velocity gradients sampled along the particle trajectory. Note that since we do not know the $\lambda$ and $\gamma$ \textit{a priori}, they must be modelled. As we will show in section \ref{results}, $\gamma$ plays an important role in balancing the gravitational accelerations \citep{skartlien_kinetic_2007}. As noted earlier, equation \eqref{closure} is derived assuming that $u_p$ has Gaussian statistics, and as we will see in the following sections, this assumption is satisfied in our flow. 

The utility of closure \eqref{closure} is that it effectively replaces the fluid velocity sampled by the particle as a sum of two terms, one proportional to the concentration gradient (the second term on the right hand side), as is usual in phenomenological models, and a drift term (the first term on the right hand side), that is proportional to the concentration itself. In the context of homogeneous turbulence, the concentration gradients will be zero, and so only the first term on the right hand size of \eqref{closure} survives. As discussed in \citet{bragg_mechanisms_2021}, this first term captures the preferential sweeping mechanism \citep{maxey_gravitational_1987,wang_dispersion_1993} that generates enhanced particle settling speeds even in homogeneous, isotropic turbulence, as observed in numerous studies such as \citet{rosa_settling_2016} and \citet{good_settling_2014}. By comparison, a model that tries to represent the vertical flux of inertial particles associated with $\varrho \langle u_p\rangle_z$ using an eddy diffusivity closure, such as that described in the next section, would never be able to represent the enhanced particle settling in the absence of gradients of the particle concentration. The closure presented in \eqref{closure} attempts to capture such enhancement in this limit. 

In their current forms, \eqref{lambda_correlation} and \eqref{gamma correlation} are still unclosed and must be modeled. There are no general analytical expressions for $\gamma$ and $\lambda$, but given certain characteristics of the turbulence, they can be estimated. The approach we will take is to assume that the turbulence is locally homogeneous (as is approximately true in the log layer of a turbulent boundary layer). There are corrections to this approach that include the influence of mean shear (see \citet{skartlien_kinetic_2007} for an example of a mean shear correction in homogeneous turbulence), but we will ignore those for simplicity. Using the locally homogeneous approximation \citep{skartlien_kinetic_2007,zhang_asymptotic_2023}, we write the following:
\begin{equation}
    \lambda \approx \langle w^2\rangle \frac{(\beta \tau_{Lp})^2}{1 + \beta \tau_{Lp}}, \label{lambda}
\end{equation}
where $\langle w^2\rangle$ is the mean squared vertical velocity of the flow, $\beta = \tau_p^{-1}$, and  $\tau_{Lp}$ is the Lagrangian timescale of the turbulence along the particle's trajectory, which can be understood as time over which the fluid motions that the particle encounters are correlated. 

It is important to note that $\tau_{Lp}$, which depends on both height and the particle's Stokes number, and $\tau$ (discussed earlier) are different quantities. \citet{bragg_particle_2012} discusses the implications and associated errors by using $\tau$ instead of $\tau_{Lp}$ when modeling $\lambda$ and $\gamma$ in a synthetic boundary layer flow. They note that in spite of the errors associated with assuming $\tau_{Lp}\approx \tau$, estimates of $\gamma$ and $\lambda$ relative to equivalent particle tracking experiments are still better than those made assuming a passive scalar approximation (which is equivalent to ignoring the drift term). Furthermore, the magnitudes of those errors tend to decrease in the interior of their domain. 

For the following analysis, we will opt to use $\tau$ in place of $\tau_{Lp}$ in \eqref{lambda}. The reason is that common models (see \citet{wang_dispersion_1993} for instance) for $\tau_{Lp}$ are strictly valid in the limit of $\mathrm{Sv}^+ \rightarrow 0$, and therefore cannot account for trajectory crossing effects (particles fall out of eddies that would be responsible for dispersing them \citep{csanady_turbulent_1963}), which arise due to the particle's Stokes settling velocity. For the following work, we use the model presented in \citet{oesterle_lagrangian_2004}, which is
\begin{equation}
    \tau = -\frac{\kappa z}{u_\tau}\frac{\langle u w\rangle}{\langle w^2\rangle}. \label{tau_def}
\end{equation}
For a simpler empirical fit to DNS data, one could alternatively use the model presented in \citet{sikovsky_particle_2019}. It is also important to note that the definition of $\tau$ is dependent only on fluid properties, thus in the one-way coupled limit, $\tau$ is not a function of the Stokes number or the settling velocity parameter. Thus, any effect on the particle transport by gravity is captured in the drift coefficient $\gamma$, discussed later.

For the purposes of this work, $\langle u w\rangle$, and $\langle w^2\rangle$ are readily available from the DNS. Therefore, we will not use any pre-existing models for these terms. However, this framework could be extended in a straightforward way: in the one-way coupled limit, both $\langle uw\rangle$ and $\langle w^2\rangle$ can be estimated by appealing to standard comprehensive wall models such as those presented in \citet{smits_highreynolds_2011,kunkel_study_2006,marusic_similarity_1997}. 

Since the total drift $-\tau_p(\gamma+\partial_z\lambda)$ captures the preferential sweeping effect, which is known to be vital for particle settling velocities in homogeneous as well as wall-bounded turbulent flows \citep{bragg_mechanisms_2021}, a model for $\gamma$ must be used. It is also important to point out that even in the absence of gravitational settling, the drift term may be non-zero due to turbulence inhomogeneity \citep{bragg_mechanisms_2021}, but for a turbulent flow where the inhomogeneity is weak, the drift term mainly captures the effect of preferential sweeping. Models for $\gamma$ do exist but they are complicated \citep{bragg_particle_2012,stafford_mass_2021}, and so for the purposes of this paper we will instead estimate $\gamma$ using 
\begin{equation}
    \tau_p\varrho\gamma \approx -\tau_p\frac{\partial}{\partial z}\lambda \varrho - \varrho \langle u_p\rangle_z, \label{residual}
\end{equation}
 with DNS data used to specify $\varrho, \langle u_p\rangle_z$ in this expression.

\subsection{Phenomenological Eddy-diffusivity approach \label{resistance}}
In large scale weather prediction models, as well as regional and global climate models, it is not feasible to represent aerosols and particles in a Lagrangian frame of reference due to computational restrictions and the astronomical number of particles found in the atmosphere. Instead, particles are often modelled as tracers in an Eulerian framework and often represent a broad size class of particles, such as those less than 10 $\mu$m (called PM$_{10}$), an approach known as a bulk scheme \citep{jenkins_wrf-chem_2022}. There are several other varieties such as sectional and modal schemes, and more information about these and their usage can be found in \citet{kukkonen_review_2012}. 

In these kinds of Eulerian frameworks, the concentration of various species follows an advection-diffusion equation according to mass conservation. In the limit of horizontal homogeneity and stationarity, mass concentration obeys:
\begin{equation}
    \frac{\partial}{\partial z}\left(-K(z)\frac{\partial C}{\partial z} - v_gC\right) = 0,
\end{equation}
where the first term in the brackets comes from assuming that the turbulent flux of the tracer can represented by an eddy diffusivity closure, and the second term represents the flux associated with the Stokes settling velocity. 
Integrating this equation with respect to height, we arrive at the following:
\begin{equation}
    -K(z)\frac{\partial C}{\partial z} - v_gC = F, \label{de}
\end{equation}
where $F$ represents the net mass flux which is constant with height. The equation can be easily solved to reveal:
\begin{equation}
    F + v_gC(z) = \left(F+ v_gC_0\right)e^{-v_g R},
\end{equation}
where $C_0$ is a reference concentration at a height $\delta_0$, and $R$ is interpreted as a resistance to transport
\begin{equation}
    R = \int_{\delta}^{z}\frac{dz'}{K(z')},
\end{equation}
which has dimensions of time over length.
As an example, to recover the familiar Rouse profile \citep{rouse_modern_1937}, the turbulent flux is assumed to exactly balance the flux due to Stokes settling, implying $F=0$, and the tracer is assumed to respond instantaneously to the eddy velocities (i.e. in the limit of vanishing inertia). This allows us to use the familiar log-layer assumption for the eddy diffusivity $K(z) = \kappa u_\tau z$ (assuming a turbulent Schmidt number of one), where $\kappa$ is the Von Karman constant. 

For net deposition conditions, $F<0$ but is constant with height. Understanding that we can write $F = -v_d(z)C(z)$ (which is identical to \eqref{phase space settling}), where $v_d$ is the deposition velocity at height $z$, we recover
\begin{equation}
    v_d = \frac{v_g\left(1 - \frac{C_0}{C(z)}e^{-v_gR}\right)}{1 - e^{-v_gR}}. \label{deposition equation}
\end{equation}
The first assumption of the above model is that the turbulent flux of particles by the turbulence can be adequately represented by the eddy-diffusivity closure
\begin{equation}
    \langle c w\rangle \approx -K(z)\frac{\partial C}{\partial z},
\end{equation}
where $c$ is the fluctuating particle concentration. As we saw in the previous section, the vertical transport for low and moderate inertial particles comes from the Stokes settling velocity plus a contribution from the preferential sampling of turbulent eddies, which, according to \eqref{closure}, is governed by the sum of a drift term proportional to the concentration, as well as a term proportional to gradient of the concentration. One of the benefits of using a closure with drift and gradient components is that in the for regions where the gradient of the mean particle concentration term becomes negligible (i.e. regions where the turbulence is nearly homogeneous and isotropic), the drift term still captures the settling enhancements for low to moderate inertia particles.

The second assumption comes from the representation of the diffusivity $K(z)$ itself, and by extension, $R$. $R$ is typically parameterized in terms how different mechanisms in the lower atmosphere affect the collection efficiency of inertial particles \citep{emerson_revisiting_2020}. These collection efficiencies are functions of the land use category in question. Several examples include forest canopies, deserts, tundras, open water, and fields \citep{farmer_dry_2021}. The primary purpose of this paper is not to modify existing parameterizations of the resistance, but to comment on how a phenomenological model compares to that of a first principles model. To that end, we use a form of the resistance which includes only the effects of turbulence and inertia. In its simplest form, we write the normalized resistance as a sum of the aerodynamic resistance and what is known as the surface resistance \citep{zhang_size-segregated_2001} (ignoring Brownian diffusion and interception, more on this below):
\begin{equation}
    u_\tau R = u_\tau R_a + u_\tau\left(\frac{\mathrm{St}^+}{\alpha + \mathrm{St}^+}\right)^{-\beta}, \label{total resistance}
\end{equation}
where $R_a$ is known as the aerodynamic resistance and takes the form
\begin{equation}
    R_a = \frac{1}{\kappa u_\tau}\log\left(\frac{z}{\delta_0}\right),
\end{equation}
while $\alpha$ and $\beta$ are positive constants that depend on the land use category \citep{zhang_size-segregated_2001}, $\delta_0$ is taken to be the maximum height of the buffer layer, and $z$ is the height at which the deposition velocity is to be calculated. In this simplified model, the second term on the right hand side of \eqref{total resistance} parameterizes the impact of inertia on the resistance. It is important to note that in operation models, there are other terms that should be included in \eqref{total resistance}. For example, for land use types without collectors (such as leaves, trees, grass) it is known that Brownian diffusion affects the deposition of particles of various sizes \citep{seinfeld_atmospheric_1998,emerson_revisiting_2020}. However, deposition enhancement by Brownian diffusion is only significant for particles much smaller than $0.1$ $\mu$m, which translates to $\mathrm{St}^+<10^{-4}$ for our DNS so it is ignored for simplicity. Furthermore, for surfaces which have collectors, their impact on the deposition must be taken into account in operational models, and this is done through a process known as interception. It turns out that the contribution to the total deposition velocity by interception is much smaller in magnitude than both the contribution by impaction and by the Stokes settling velocity anyways \cite{farmer_dry_2021}, and is therefore ignored in \eqref{total resistance}.  Thus, the simplest model to compare the DNS to is simply to include a model for impaction (i.e. the second term on the right hand side of \eqref{total resistance}). In the limit of $\mathrm{St^+}\rightarrow 0$, the second term above takes a form proportional to $(\mathrm{St}^+)^{-\beta}$ indicating that $\mathrm{R}\rightarrow \infty$. In this limit, the model shown in \eqref{deposition equation} says that $v_d \rightarrow v_g$ which is what is expected in the limit of small $\mathrm{St}^+$. In the limit of $\mathrm{St^+}\rightarrow \infty$, the overall resistance at a height $z$ approaches a constant value. According to \eqref{deposition equation}, since $v_g$ is coupled to $\mathrm{St}^+$ through the constancy of $g$, this implies that again $v_d \rightarrow v_g$, which is the correct relationship in the ballistic limit.

Thus, in summary, the two models we are comparing yield the following forms for the settling velocity
\begin{eqnarray}
    &v_d(z) = \frac{v_g}{1 - e^{-v_gR(z)}}, \label{deposition} \\
    &\langle w_p \rangle_z  = -\tau_pg -\tau_p\gamma - \frac{\tau_p}{\varrho}\frac{\partial}{\partial z}\left(\lambda \varrho - \varrho S\right). \label{settling}
\end{eqnarray}
Note that in \eqref{deposition}, we have ignored the contribution from $C_0/C(z)$ in \eqref{deposition equation}. As this term approaches unity (the concentration at a measurement height $z$ is nearly identical to the reference concentration), it implies that there is no vertical variation in the concentration profile, and mathematically, $v_d$ relaxes back to $v_g$. This result is consistent within the context of this model, as the gradient-diffusion closure implicitly assumes that all turbulent transport must be a consequence of gradients of the mean concentration. Of course this is certainly not true of real turbulent flows, e.g., homogeneous isotropic turbulence. On the other hand, as $C_0/C(z)$ approaches zero, (the concentration at the measuring height $z$ is much larger than the reference concentration), we approach a state fully enhanced by turbulence, and is therefore bounded from above by \eqref{deposition}. Thus, \eqref{deposition} represents the upper bound on the deposition velocity irrespective of the concentration profile, and this is what we will compare the results from the phase space modelling to. Note that \citet{hoppel_particle_2005} incorrectly sets $C_0$ to zero to arrive at \eqref{deposition}.

As a final point, in this work, we will use ``settling velocity'' and ``deposition velocity'' interchangeably, but in an effort to discriminate between the two approaches discussed in the following results, we will use $\langle w_p\rangle_z$ when referring to the settling velocity computed using the phase space approach, and we will use $v_d$ when discussing the deposition velocity associated with the phenomenological modeling approach. Furthermore, since the particles do not change their radius over the course of the simulation, the number concentration, denoted by $\varrho$, is linearly related to the mass concentration, denoted by $C$. Therefore, the number flux is also linearly related to the mass flux. 

\section{Results \label{results}}

\subsection{Settling velocity enhancement for Numerical Experiment 1}
\begin{figure}
    \centering
    \includegraphics[width=\textwidth]{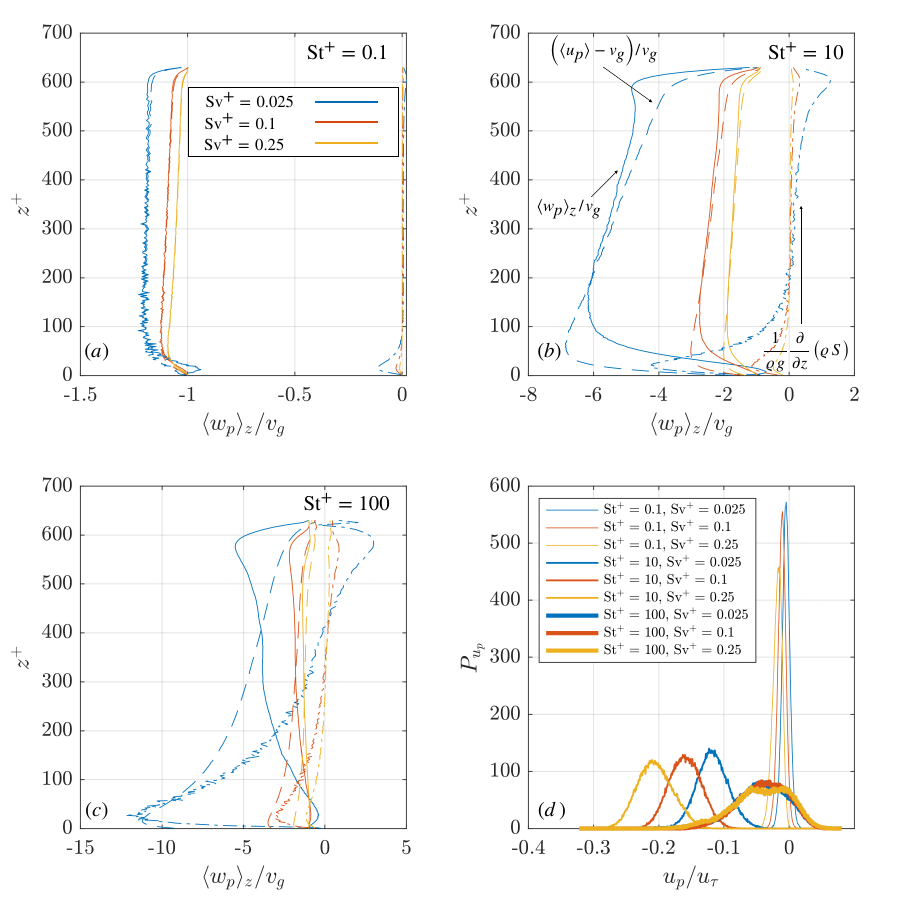}
    \caption{Profiles of the averaged settling velocity (dashed curves), the averaged fluid velocity sampled by the particle (solid curves), and the averaged diffusive flux and turbophoretic drift terms (dot-dashed curves) for $\mathrm{St}^+ = 0.1$, 10, and 100 in panels (a-c) respectively. Information for the cases shown can be found in table \ref{tab:cases}. Recall that the diffusive flux and the turbophoretic drift terms are grouped together. Panel (d) shows PDFs of the vertical fluid velocities sampled by the particles normalized by $u_\tau$ in the interior by the particles over a given simulation. Cases with $\mathrm{Sv}^+$ of 0.025, 0.1, and 0.25 are shown in blue, red and yellow curves respectively, while the thickness of the curves increases as $\mathrm{St}^+$ is increased.}
    \label{fig:mass_flux_and_sampling_dist}
\end{figure}

We will begin the analysis and comparison by briefly highlighting the main characteristics for the cases in Numerical Experiment 1. 
Figure \ref{fig:mass_flux_and_sampling_dist}(a-c) show the DNS-computed settling velocity (dashed curves), the sum of the average sampled velocities, and the Stokes settling velocity (solid curve) and the diffusive flux and turbophoretic term (dot-dashed curves) for $\mathrm{St}^+=0.1$, 10 and 100 respectively and for all three values of $\mathrm{Sv}^+$ for each. The third term on the right hand side of \eqref{settling evolution} has been omitted as it is known to be negligible. First, we will focus on the cases with $\mathrm{St}^+ = 10$, figure \ref{fig:mass_flux_and_sampling_dist}(b), as it clearly shows the role of each settling mechanism. We can identify that the overall settling enhancement relative to the Stokes settling velocity decreases as $\mathrm{Sv}^+$ is increased. Within the interior of the domain, (about $100<z^+<600$) the grouped diffusive flux and the turbophoretic drift term is negligible within the interior for all cases except the case with the lowest $\mathrm{Sv}^+$, which begins to increase with height implying an upward tendency, though it is weak. Though these terms are implicitly dependent on the particle velocities (and thus $\mathrm{Sv}^+$), they are explicitly dependent on $\mathrm{St}^+$, so by increasing $\mathrm{Sv}^+$, the relative contribution in the boundary layer decreases. Therefore, we can see that the decreased turbulent settling velocity must come from the reduction in the difference between the sampled velocities and the Stokes settling velocities as $\mathrm{Sv}^+$ is increased.

For $\mathrm{St}^+ = 0.1$, figure \ref{fig:mass_flux_and_sampling_dist}(a), the settling velocity in the interior is well approximated by the sum of the Stokes settling velocity and the average velocity sampled by the particles, as this was the dominant balance at low Stokes number, pointed out by \cite{bragg_mechanisms_2021} since all other terms in \eqref{settling evolution} are proportional to $\mathrm{St}^+$. By comparison, adequate representation of the average settling velocity for $\mathrm{St}^+=100$, figure \ref{fig:mass_flux_and_sampling_dist}(c), requires the inclusion of all terms in \eqref{settling evolution} (aside from the term including $\langle w_p\rangle^2$). Thus, to fully model the settling velocity at $\mathrm{St}^+=100$ at the current $\mathrm{Re}_\tau$, a separate model of the diffusive flux and the turbophoretic drift term (or equivalently of the turbophoresis and the particle fluctuating covariance tensor) is required. Since the goal of this work is to compare models of the sampled velocities ($\langle u_p\rangle_z$), we will not comment on models of these quantities, but refer the reader to \citet{zhang_asymptotic_2023} for a detailed discussion. 

We can see by consulting figure \ref{fig:mass_flux_and_sampling_dist}(d), which shows normalized distributions of the sampled velocities in the range $40< z^+ < 400$ (outside of the buffer layer and away from the upper boundary), that for a given low or moderate $\mathrm{St}^+$ (0.1 or 10, denoted by thin and medium line thickness respectively), the magnitude of the sampled velocity relative to $u_\tau$ increases with $\mathrm{Sv}^+$ (each $\mathrm{Sv}^+$ is denoted by a different line color).  Comparing the different representations of the particle sampling in this way highlights the complex behaviour of the preferential sampling. For example, taken together figure \ref{fig:mass_flux_and_sampling_dist}(b) and (d) (medium thickness curves) highlight that the magnitude of the sampled velocities increases sub-linearly with $\mathrm{Sv}^+$. For example, this means that to double $\mathrm{Sv}^+$ doubles $v_g$, but increases the magnitude of $\langle u_p\rangle_z$ by less than a factor of two. Thus, the overall settling velocity should increase sub-linearly with $\mathrm{Sv}^+$.
This is in contrast with the distribution of velocities sampled by particles with $\mathrm{St}^+ =100$ which appear to show independence of $\mathrm{Sv}^+$ (i.e. the distributions are qualitatively similar regardless of the choice of $\mathrm{Sv}^+$).

\begin{figure}
    \centering
    \includegraphics[width=\textwidth]{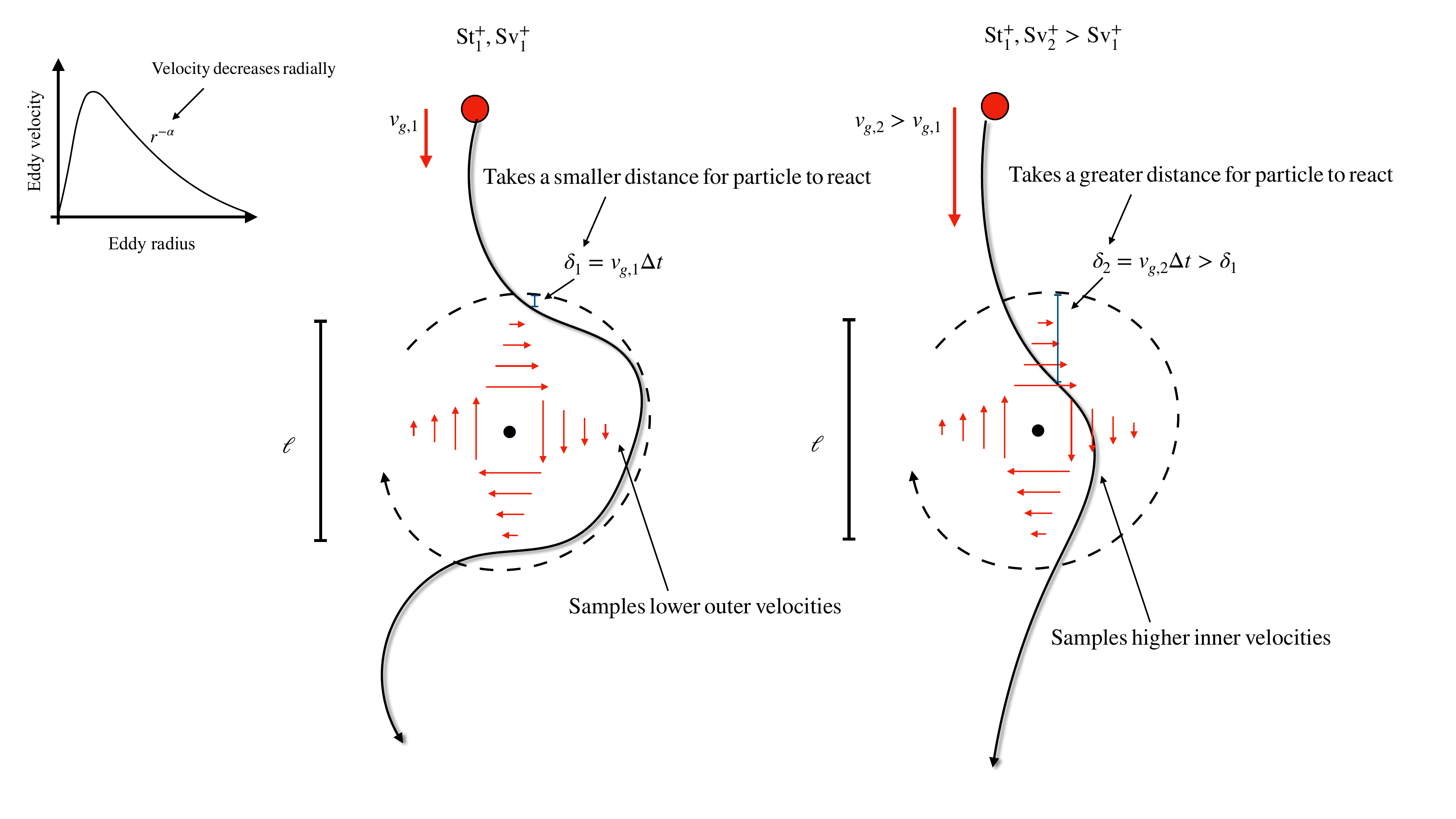}
    \caption{The modification of preferential sweeping due to the variation in $\mathrm{Sv}^+$ at constant $\mathrm{St}^+$. The top left corner is a prototypical relationship between the velocity of a turbulent eddy and the radial distance away from its center. The middle and rightmost panel show a turbulent eddy of size $\ell$ (outline by the dashed curve), and the qualitative velocity distribution denoted by red arrows within. The particle (the red circle) follows the trajectory denoted by the solid curved arrow. For a given low to moderate Stokes number and $\mathrm{Sv}^+$, the particle penetrates a certain distance into the eddy before as it relaxes to the local velocity field. It is then swept to the downward side of the eddy. If $\mathrm{Sv}^+$ is increased (or alternatively $g$, then the particle penetrates further into the eddy as it is settling at a higher rate. Therefore, it samples larger inner velocities.}
    \label{fig:Sv_sample_schematic}
\end{figure}

These distributions imply that the traditional view of preferential sweeping from \citet{wang_settling_1993} is modified as $\mathrm{Sv}^+$ is varied independently from $\mathrm{St}^+$. The modification to the traditional picture of preferential sweeping for a given $\mathrm{St}^+$ due to $\mathrm{Sv}^+$ is demonstrated conceptually in figure \ref{fig:Sv_sample_schematic}.

As a settling inertial particle approaches an eddy, the particle is preferentially swept to the downward side of the eddy, meaning that inertial particles will preferentially sample negative velocities. For a fixed $\mathrm{St}^+$, increasing $\mathrm{Sv}^+$ implies an increase in the Stokes settling velocity. This means that in one relaxation timescale, a particle travels a distance (normalized by wall units) $\delta \approx g\tau_p^2u_\tau/\nu$. This distance is only approximate because the speed at which the particle enters a given eddy is actually a function of its history, as it was presumably accelerated by eddies that it had previously encountered. Thus, for a fixed $\mathrm{St}^+$, increasing $\mathrm{Sv}^+$ allows the inertial particle to penetrate deeper into a given eddy before it can relax to the local flow conditions.
Though the exact form of a turbulent eddy is not known, many models \citep{pullin_vortex_1998} point to turbulent eddies having velocity distributions that are inversely proportional to the distance from the eddy core. Thus, a particle penetrating further into a turbulent eddy samples higher velocities before it is relaxes to the local fluid velocity. This means that, on average, particles at a given Stokes number sample higher velocities at higher $\mathrm{Sv}^+$. 
This picture is true until the $\mathrm{St}^+$ is large enough that the particle trajectory becomes uncorrelated from the fluid motion regardless of $\delta$.  

Particle-turbulence interactions in a real turbulent flow are of course much more complex than those in this simple cartoon. Nevertheless, it provides some insights into how the sampling of the flow by the particles could increase with increasing $\mathrm{Sv}^+$ for a given $\mathrm{St}^+$. It cannot be true in general, however, because for $\mathrm{Sv}^+\to\infty$ the particles settle ballistically through the eddies and there is no preferential sweeping. 

\subsection{Modeling the fluid velocity sampled by the particle \label{pdf approach}}

\begin{figure}
    \centering
    \includegraphics[width=\textwidth]{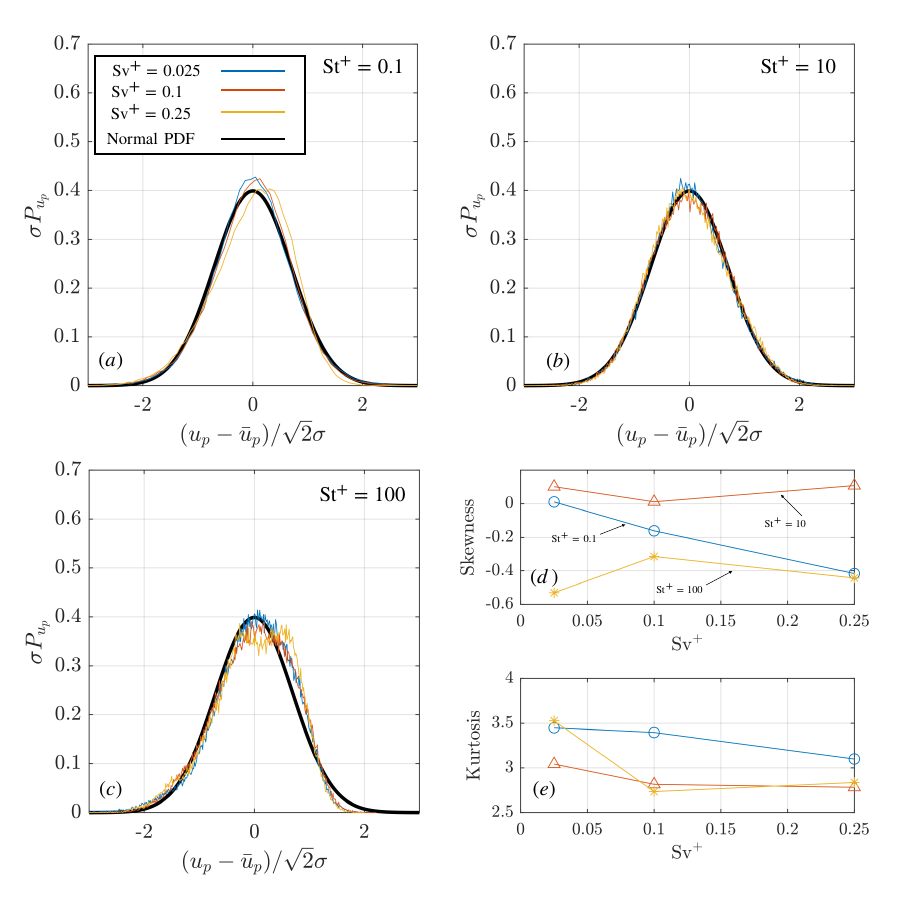}
    \caption{PDFs of the sampled velocities for all cases in Numerical Experiment 1 plotted in standard form and compared to a normal distribution (black curves). Panels (a-c) show data for $\mathrm{St}^+ = 0.1$, 10, and 100, which curve color corresponds to $\mathrm{Sv}^+$. Panels (d) and (e) show the skewness and kurtosis of the data in panels (a-c).}
    \label{fig:standard_pdf}
\end{figure}

In the following sections, we use a combination of the DNS data and the model in \eqref{lambda} (with $\tau$ in place of $\tau_{Lp}$) to estimate the magnitudes of the terms in \eqref{closure} as well as their behaviour with respect to $\mathrm{St}^+$ and $\mathrm{Sv}^+$. Recall that the base assumption regarding the applicability of \eqref{closure} is that the vertical component of the sampled velocities are normally distributed about their mean value \citep{skartlien_kinetic_2007}. We show this qualitatively and quantitatively in figure \ref{fig:standard_pdf}, where we compare the PDFs shown in figure \ref{fig:mass_flux_and_sampling_dist}(d) to a standard normal distribution. We can see that for low and moderate $\mathrm{St}^+$ (figs \ref{fig:standard_pdf}(a) and (b)) that the distributions are qualitatively Gaussian, but there is weak non-Gaussianity for all cases, quantitatively confirmed by the skewness and kurtosis shown in panels (d) and (e) (orange and blue curves). For $\mathrm{St}^+ =100$, figure \ref{fig:standard_pdf}(c), the sampled velocities within the interior are more complex, and this is reflected in their negative skewness (right-skewed) values, but comparable kurtosis (note that a standard normal distribution has a kurtosis of 3.0). Finally, it is interesting to note that the cases with $\mathrm{St}^+ =10$ exhibit the weakest non-gaussianity in the sampled velocities.  

\begin{figure}[htbp!]
    \centering
    \includegraphics[width=\textwidth]{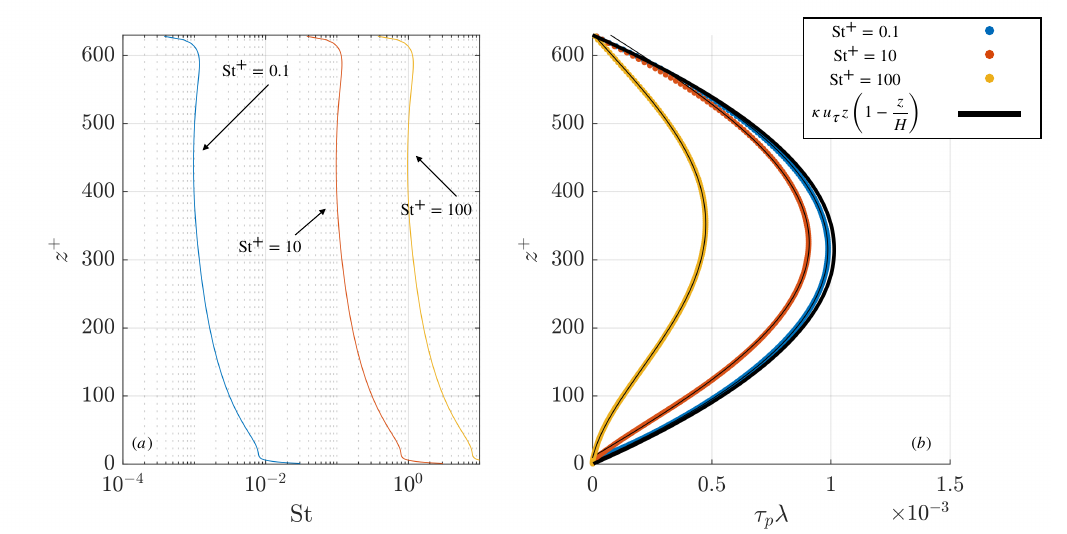}
    \caption{Panel (a) shows the integral Stokes number for $\mathrm{St}^+ = 0.1,$ 10, and 100 and panel (b) shows the computed $\tau_p\lambda$ for each $\mathrm{St}^+$ with $\mathrm{Sv}^+ = 0.1$. (dots), and a fourth order polynomial fit to each dataset (thin black line) and the quadratic diffusivity profile using a log layer scaling from \citet{hoppel_particle_2005} (thick black line).}
    \label{fig:integral_Stokes}
\end{figure}

Next, using the DNS data, we can calculate $\tau$ using \eqref{tau_def} to determine both the integral Stokes number, $\mathrm{St}$, as well as $\tau_p\lambda$. These quantities are shown in figure \ref{fig:integral_Stokes}(a) and (b) respectively. We can see that across the interior of the domain (say in the range $50<z^+<500$), the range of the integral Stokes number for a given $\mathrm{St}^+$ is less than an order of magnitude. Recall that since $\tau$ is only a function of fluid properties, cases with fixed $\mathrm{St}^+$ and different $\mathrm{Sv}^+$ will have the same integral Stokes number. Furthermore, we can see that by increasing $\mathrm{St}^+$ by a factor of 10 or 100, we also increase $\mathrm{St}$ at a given height by a factor of 10 or 100.

Note that for small and  moderate $\mathrm{St}^+$ (0.1 or 10), the integral Stokes number is less than unity across the entire interior of the domain, suggesting only a weak decoupling of the particle trajectories from the Lagrangian fluid trajectories there. This correlates with the fact that the distributions of sampled velocities are approximately Gaussian (see figure \ref{fig:standard_pdf} (a-b)), so we should expect the closure discussed in \eqref{closure} to be approximately valid for these cases. For $\mathrm{St}^+=100$, the integral Stokes number varies from about one near the midpoint of the domain to five near the viscous sublayer, possibly reflecting the stronger skew of the sampled velocities for this value of $\mathrm{St}^+$.

Given our model for $\tau$, we can now use $\langle w^2 \rangle$ to compute $\tau_p\lambda$, which can be interpreted as the diffusivity for the ``gradient" based part of the closure. Shown in \ref{fig:integral_Stokes}(b) are values of $\tau_p\lambda$ computed by the DNS for all cases at $\mathrm{Sv}^+=0.1$ in Numerical Experiment 1, and these are plotted as circular markers. A fourth order polynomial fit in the region $40 < z^+ 600$ is overlaid on the data. This interval was chosen to avoid non-physical boundary effects. Recall that since $\lambda$ is only a function of fluid properties ($\tau$ and $\langle w^2\rangle$), as well as the relaxation timescale ($\tau_p$), changes in $\mathrm{Sv}^+$ due to changes in $g$ will not change the profile of $\lambda$. 

The choice to use a fourth order model is arbitrary, but qualitatively speaking, other common forms of the scalar and momentum diffusivities in phenomenological closures often take similar even order polynomial forms. For example, for a neutrally stratified turbulent boundary layer with a logarithmic mean velocity profile, a quadratic model for the momentum diffusivity can be easily derived to be $K_q(z) = \kappa u_\tau z\left(1 - \frac{z}{H}\right)$ \citep{fischer_longitudinal_1973,hoppel_particle_2005}. This profile is plotted on figure \ref{fig:integral_Stokes}(b) as a thick black curve. Furthermore, for an unstable planetary boundary layer, scalar diffusivities are often represented by polynomials of fractional order in the vertical coordinate \citep{holtslag_eddy_1991,wyngaard_top-down_1984}, and a similar approach was taken by \citet{nissanka_parameterized_2018} who used a blend of linear and cubic scalar diffusivity profiles to model particle dispersion within the marine atmospheric boundary layer. Often, for the bottom 100 meters of the ABL, it is common to take a diffusivity which is linear in height (see \cite{kantha_small_2000} for example).
We can see from this plot is that $\tau_p\lambda$ for $\mathrm{St}^+=0.1$ approaches the diffusivity profile predicted by a quadratic log layer scaling. This is a consequence of the particle acting as passive scalars in the low $\mathrm{St}^+$ limit. It is known in that in this limit, the drift coefficient tends to take the form
\begin{equation}
    \gamma = -\frac{\partial \lambda}{\partial z},
\end{equation} thus \eqref{closure} only contains the gradient diffusion term. This is known as the passive scalar approximation \citep{bragg_particle_2012}. Further increasing $\mathrm{St}^+$ to 10 or 100, we decrease the diffusivity relative to the theoretical functional form. According to \eqref{lambda_correlation} (the definition of $\lambda)$, this must come from the fact that the vertical displacements of the particles associated with the sampled velocities become smaller. This can also be understood as a consequence of the crossing trajectory effect discussed in \citet{csanady_turbulent_1963} where a reduction in the vertical diffusivity arises from the particles increasing settling velocity through an increase in $\tau_p$.

\begin{figure}[htbp!]
    \centering
    \includegraphics[width=\textwidth]{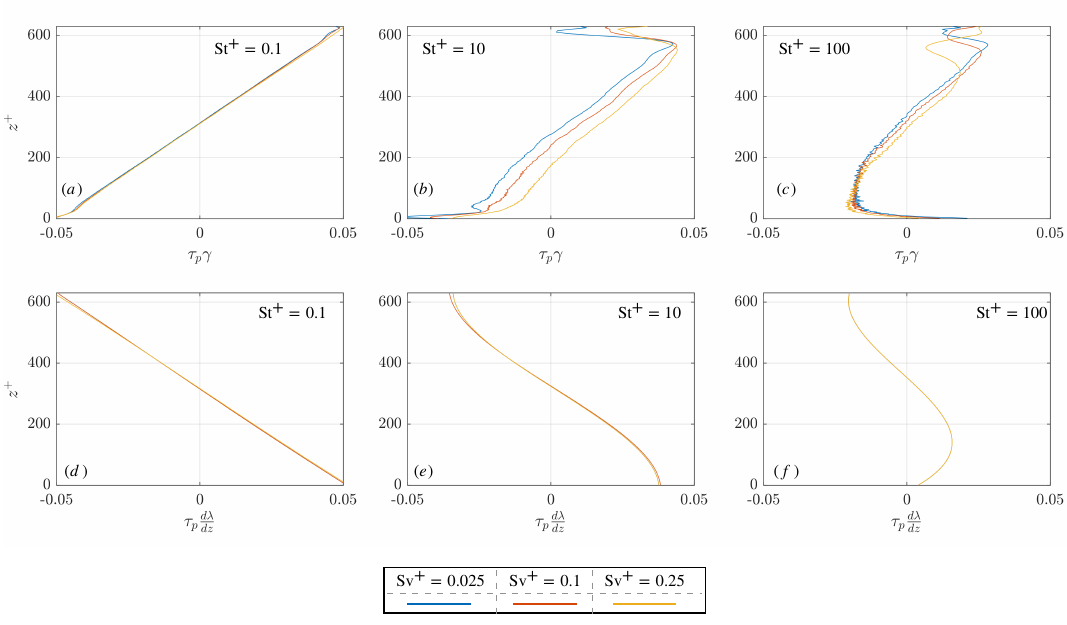}
    \caption{A comparison of the magnitude of terms that make up the drift coefficient. Profiles of $\tau_p\gamma$ are presented in panels (a--c) and profiles of $\tau_p\frac{\partial \lambda}{\partial z}$ are presented in panels (d--f).  Each curve within a panel corresponds to a different $\mathrm{Sv}^+$, while each column corresponds to a different $\mathrm{St}^+$.}
    \label{fig:gamma_lambdaz_compare}
\end{figure}

Using the estimate for $\tau_p\lambda$ from figure \ref{fig:integral_Stokes}(b), the known number concentration $\varrho$, and averaged sampled velocities $\langle u_p\rangle_z$ from the DNS, we can estimate the residual from Eq. \eqref{residual} to help us to understand what $\gamma$ ought to be for Eq. \eqref{closure} to adequately represent the sampled velocities. Furthermore, we can compare this to $\frac{\partial \lambda}{\partial z}$ to identify the dominant contribution to the drift term. These quantities are plotted in figure \ref{fig:gamma_lambdaz_compare}. 

By virtue of the the fact that $\tau$ is a fluid property, the definition of $\lambda$ we use in this work is not a function of $\mathrm{Sv}^+$. Thus, dependence on $\mathrm{Sv}^+$ should be encapsulated within $\gamma$.
We can see that for $\mathrm{St}^+=0.1$ in figures \ref{fig:gamma_lambdaz_compare}(a) and (d) that there is negligible $\mathrm{Sv}^+$ dependence in both terms, as the curves for each value of $\mathrm{Sv}^+$ are nearly coincident. Furthermore, these results are consistent with the passive scalar approximation discussed earlier ($\gamma = -\frac{\partial \lambda}{\partial z}$), which is appropriate in the limit $\mathrm{St}\rightarrow 0$.
By increasing $\mathrm{St}^+$ to 10, profiles of $\tau_p\gamma$ clearly depend on $\mathrm{Sv}^+$ within the interior. Qualitatively speaking, it appears that the profiles within the interior (away from the boundaries) are of similar slope but are offset by a constant dependant on $\mathrm{Sv}^+$. This means that to adequately model enhanced settling, inclusion of the drift coefficient $\gamma$ as well as its dependence on $\mathrm{Sv}^+$ in a model is imperative, especially for moderate Stokes number. Lastly, we can see that the behaviour of the $\gamma$ profiles becomes more complex for $\mathrm{St}^+ = 100$. There are some differences in the $\gamma$ profiles due to $\mathrm{Sv}^+$, but they are only clear in the range $200 < z^+ < 400$, with the $\gamma$ profiles collapsing in the lower third of the domain, and potentially being affected by the boundary conditions near the upper boundary.

Furthermore, the positive regions of profiles of $\gamma$ within the interior of the domain indicate that downward changes in the particle vertical locations are correlated with positive local gradients of the sampled fluid velocities, indicated by Eq. \eqref{gamma correlation}. By decreasing $\mathrm{Sv^+}$, we increase the thickness of the near boundary region where $\gamma<0$ meaning that vertical changes in the particle locations are instead anti-correlated with gradients of the sampled velocities. We can see from Fig. \ref{fig:mass_flux_and_sampling_dist}(b) that this problem comes from the fact that the gradient of the sampled velocities switches sign in this region.

\begin{figure}[htbp!]
    \centering
    \includegraphics[width=\textwidth]{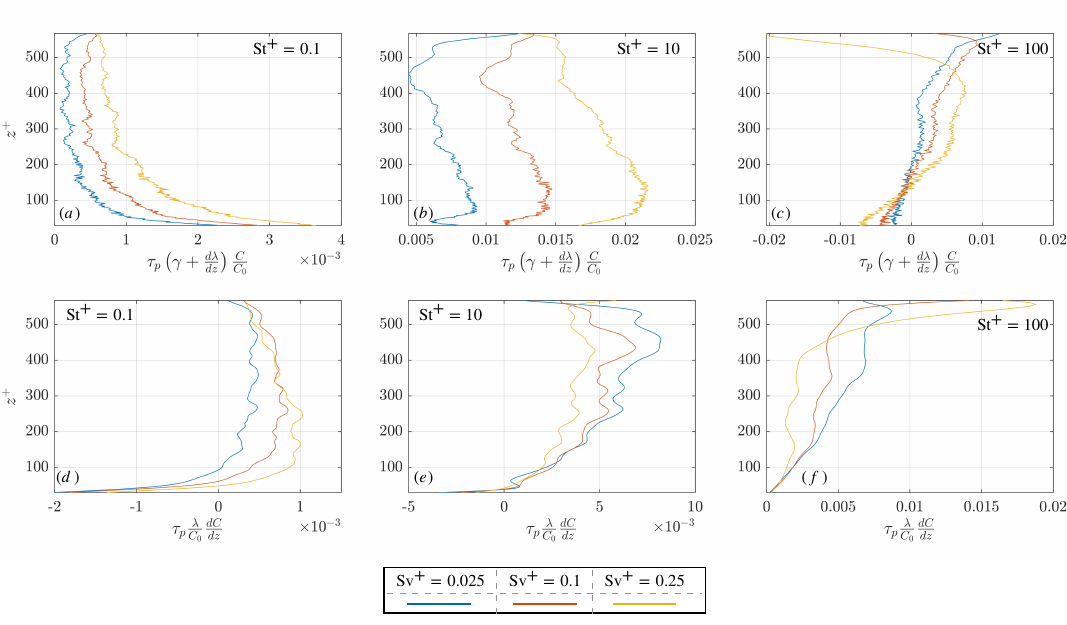}
    \caption{Profiles of the drift component of \eqref{closure} are presented in panels (a--c) and profiles of the diffusive component of \eqref{closure} are presented in panels (d--f).  Each curve within a panel corresponds to a different $\mathrm{Sv}^+$, while each column corresponds to a different $\mathrm{St}^+$. Data smoothing has been applied to the concentration profiles to in order to smooth the derivatives.}
    \label{fig:closure_components}
\end{figure}

Next, with detailed information about $\lambda$ and $\gamma$ in hand, the drift and diffusion components of Eq. \eqref{closure} are shown in figure \ref{fig:closure_components}. The drift components are shown in Fig. \ref{fig:closure_components}(a--c) and the diffusion components are shown in Fig. \ref{fig:closure_components}(d--f). Each column corresponds to a value of $\mathrm{St}^+$ while each curve corresponds to a specific value of $\mathrm{Sv}^+$. 

The primary finding here is that the drift contribution implied by the DNS is in fact up to an order of magnitude more important than the diffusion component for moderate $\mathrm{St}^+$ and high $\mathrm{Sv}^+$, shown in figures \ref{fig:closure_components}(b) and (e), and this must come from the enhancement by preferential sweeping. This difference decreases as $\mathrm{Sv}^+$ is decreased since the diffusion component tends to increase in magnitude overall, probably due to the crossing trajectory effect, while the drift coefficient decreases in magnitude overall. It is sensible for the drift coefficient to decrease in magnitude in the limit of $\mathrm{Sv}^+\rightarrow 0$ since in this limit, there is no net downward flux and on average particles begin to sample upward motions as frequently as downward motions within the interior. 
On the other hand, for both high and low $\mathrm{St}^+$, the drift component and the diffusion components are at least comparable in magnitude within the interior of the domain. 

Roughly speaking, we can see that as the Stokes number increases, the magnitude of the diffusive profiles within the interior for all $\mathrm{Sv}^+$ increases by a factor of 3 to 10 (depending on $\mathrm{Sv}^+$) between $\mathrm{St}^+=0.1$ and 10, while the differences between the diffusive profiles for $\mathrm{St}^+=10$ and $\mathrm{St}^+=100$ are not nearly as large. On the other hand, the overall magnitude of the drift coefficient reaches a peak at the moderate Stokes number. It is interesting to note that in figure \ref{fig:closure_components}(d), the diffusion for the $\mathrm{Sv}^+=0.025$ case actually has the lowest diffusion magnitude and the lowest drift, implying that the averaged sampled velocities begin to approach zero in the passive scalar limit. 

These results indicate that an understanding of the parametric behaviour of the drift correction is imperative for adequate representation of the fluid velocities sampled by the particle in the log layer for low and moderate Stokes number. These are important observations since for phenomenological modeling efforts, the entirety of the turbulent flux is assumed to be encoded in an eddy diffusivity closure, but it is clear that in the moderate $\mathrm{St}^+$ limit, the majority of the settling velocity enhancement actually comes from the drift component, and the magnitude and structure of the drift strongly depends on both $\mathrm{St}^+$ and $\mathrm{Sv}^+$.

\subsection{Deposition velocity according to the phenomenological approach}
Prior to the usage of the closure for the sampled vertical fluid velocities, the PDF-based approach is exact, in the sense that it introduces no additional assumptions about either the particle or fluid dynamics aside from those already in the equations of motion themselves \citep{bragg_mechanisms_2021}. However, it is very general and identifying opportunities for modeling simplification can be quite challenging for weather and dispersion applications. Therefore, it is useful to identify how the DNS data (from which we have provided estimates for $\gamma$ and $\lambda$) compares with phenomenological models which are used ubiquitously throughout the atmospheric dispersion literature. To do this, we will compare the DNS data with the model presented in Eq. \eqref{deposition equation}. 

\begin{figure}[htbp!]
    \centering
    \includegraphics[width=\textwidth]{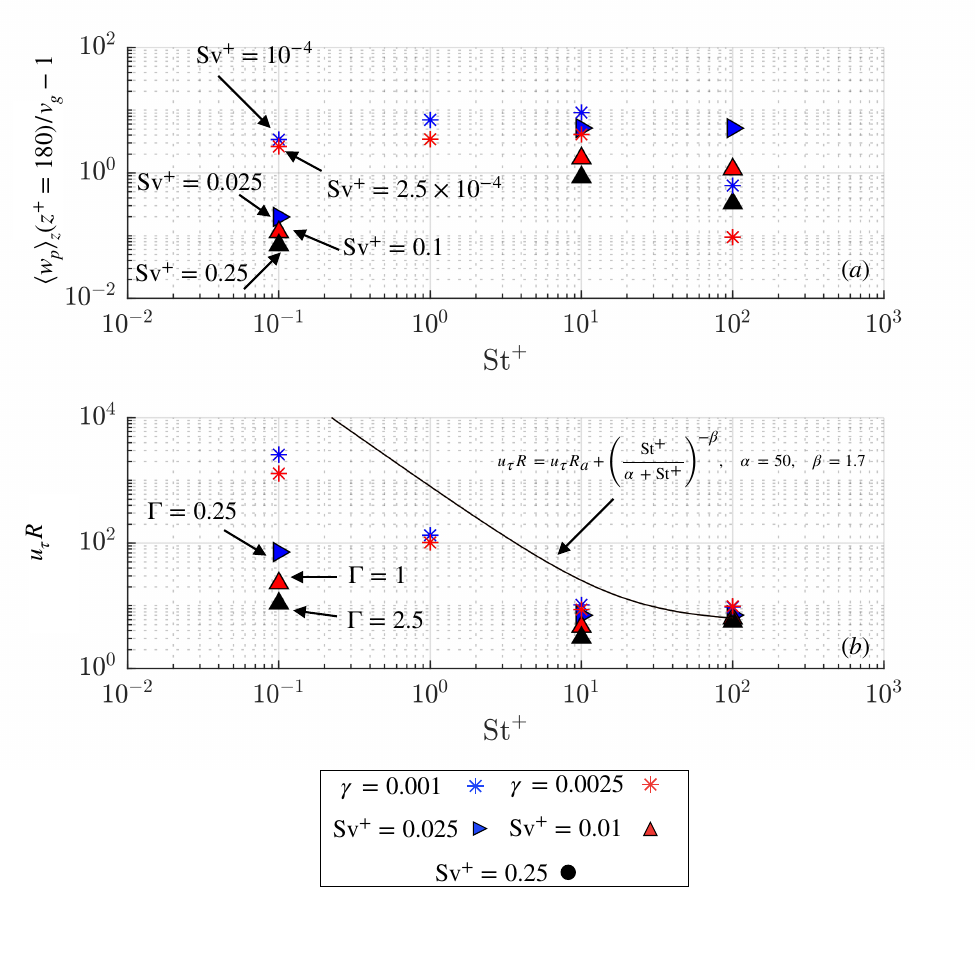}
    \caption{The settling velocity enhancement plotted against $\mathrm{St}^+$ at $z^+ = 180$, panel (a), and the normalized resistance computed from the DNS plotted against $\mathrm{St}^+$ at $z^+ = 180$, panel (b). Data from cases in table \ref{tab:cases} are plotted as filled markers. Data from cases in table \ref{tab:cases_slinn} are plotted as stars. The resistance model from \cite{zhang_size-segregated_2001} with $\alpha = 50$ and $\beta = 1.7$ is included on panel (b).}
    \label{fig:settling_enhancement_and_resistance}
\end{figure}

Figure \ref{fig:settling_enhancement_and_resistance}(a) shows the settling velocity enhancement for Numerical Experiment 1 as filled markers, and the settling velocity enhancement of Numerical Experiment 2 as star markers. The common behaviour is that there is a maximum in the settling enhancement somewhere in the neighborhood of $\mathrm{St}^+ = 10$ ($\mathrm{St}$ in the range of 0.1 to 1 according to figure \ref{fig:integral_Stokes}) for all cases, which qualitatively agrees with past studies on the settling velocity enhancement of inertial particles \citep{wang_settling_1993,rosa_settling_2016}. However, at constant $\Gamma$ (stars), the enhancement is sustained at very low $\mathrm{St}^+$, suggesting that the velocities sampled by the particles tend to become more important in determining the enhancement at such low $\mathrm{Sv}^+$. It has been shown previously that for $\mathrm{Sv}^+$ in the neighborhood of $10^{-3}$ or $10^{-4}$, that the magnitude of $\mathrm{Sv}^+$ does not play a significant role in the dynamics of the particles outside a very small region near the bottom boundary \citep{bragg_settling_2021}, so the dominant effect leading to settling enhancement is the preferential sweeping by turbulent eddies. In these cases, distributions of the sampled velocities tend to converge around the same value when normalized by the Stokes settling velocity (not shown).

Figure \ref{fig:settling_enhancement_and_resistance}(b) shows the resistance computed from the DNS for all cases in experiments 1 and 2 as well as the theoretical curve from \citet{zhang_size-segregated_2001} using $\alpha = 50$ and $\beta = 1.7$. We can see that as $\Gamma$ becomes smaller, the data begins to approach this empirical curve, whereas when $\Gamma$ becomes large, it is clear that the data diverges from it. Different models for the inertial resistance such as those from \citet{slinn_predictions_1980,giorgi_particle_1986,zhang_size-segregated_2001,emerson_revisiting_2020} will change the slope of this curve, but the point is that as far as we can tell, none of these models attempt to define the variation in slope as a function of $\Gamma$, even though it so clearly matters. Furthermore, as the Stokes number increases, the data points tend to group together and approach the asymptotic aerodynamic resistance component. In the model from \citet{hoppel_particle_2005}, this corresponds to the resistance acquiring a constant value, leading to negligible velocity enhancement as $\mathrm{St}^+$ increases, which our DNS qualitatively agrees with.

An important point to mention is that the empirically determined resistance was designed to be valid for geophysically relevant values of $\Gamma$ in the locations where the data was taken. For example, under normal atmospheric boundary layer conditions on earth, we expect $u_\tau$ to take values in the neighborhood of 10 $\mathrm{cm/s}$ to 1 $\mathrm{m/s}$ (see \citet{vickers_formulation_2015} for example), meaning $\Gamma$ takes on values between $10^{-4}$ and $10^{-1}$. We can see that when $\Gamma$ is close to this estimated range, the data tends to approach the empirical curve, while for $\Gamma$ out of this range, the data appears to diverge from this relationship. This means that we should not necessarily expect the empirical curve to match the DNS data quantitatively for the experiments in table \ref{tab:cases}, or more generally, environments where $\Gamma$ significantly diverges from the geophysically relevant range. 

\begin{figure}[htbp!]
    \centering
    \includegraphics[width=\textwidth]{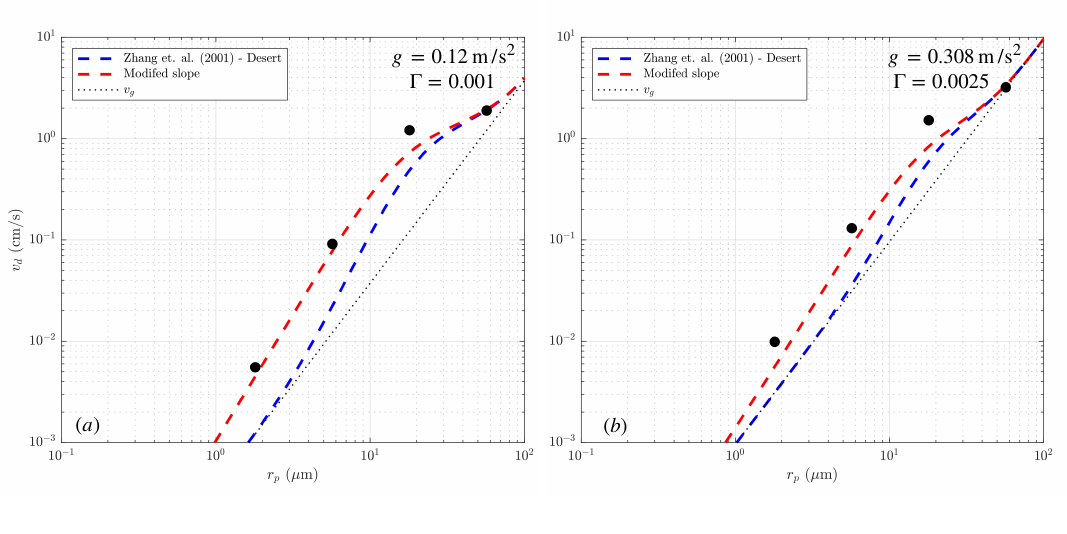}
    \caption{Data from table \ref{tab:cases_slinn} re-dimensionalized and plotted as black markers to match with standard deposition curves from \cite{hoppel_particle_2005}. The Stokes settling velocity as a function of particle radius is plotted by the dotted line. The parameterization presented in \cite{zhang_size-segregated_2001} using $\beta = 1.7$ and $\alpha = 50$ (corresponding to a desert or tundra setting) is plotted by the blue dashed line. A similar parameterization with $\beta = 1.3$ and $\alpha =50$ is shown by the red dashed curve. Panel (a) shows the case with $\Gamma = 0.001$ and panel (b) shows the case with $\Gamma = 0.0025$.}
    \label{fig:hoppel_comparison}
\end{figure}

The implications of the under-estimate of the overall resistance is that for larger $\Gamma$, over-representing the resistance will lead to an under-estimate of the deposition velocity for low and moderate Stokes number. A demonstration of this is shown in figure \ref{fig:hoppel_comparison} for the re-dimenisonalized data presented in table \ref{tab:cases_slinn}. Cases with $\Gamma = 0.001$ are shown in figure \ref{fig:hoppel_comparison}(a) while cases with $\Gamma = 0.0025$ are shown in figure \ref{fig:hoppel_comparison}(b). To determine an effective particle radius, we assume that the density for all particles is that of quartz sand ($\rho_p = 2650$ $\mathrm{kg/m^3}$) allowing us to calculate the equivalent particle radius based on the particle's $\mathrm{St}^+$. On these plots, we have plotted the Stokes settling velocity as a function of particle radius (the dotted lines), the deposition velocity in Eq. \eqref{deposition equation} using the resistance from \cite{zhang_size-segregated_2001} with $\alpha = 50$ and $\beta = 1.7$ (blue dashed lines), as well as a modified resistance where we have arbitrarily adjusted $\beta$ to be 1.3 (red dashed curves). Thus, it is clear that the deposition model in Eq. \eqref{deposition equation}, while practically useful, requires calibration data from the environment one wishes to study in order to be effective, and if existing parameterizations are used in the incorrect context, we can see that this leads to serious errors, some cases being as high as 500\%. 

Furthermore, we can see that under the \citet{zhang_size-segregated_2001} resistance model, the maximally enhanced particles at $\mathrm{St}^+ = 10$ ($r_p \approx 10-11$ $\mu$m for our DNS) is also under-estimated, but it is important to point out that it is due to a different reason than that discussed above. According to the primary assumption that the resistances are added together, it is clear that their minimum value is the aerodynamic resistance $R_a$. However, for large $\Gamma$ and $\mathrm{St}^+=10$, the computed resistance is actually smaller than $R_a$, meaning that the enhanced settling velocity will always be under-estimated for moderate $\mathrm{St}^+$ regardless of the choice of $\alpha$ and $\beta$. This means that the resistance-based parameterization in its current form is inadequate to describe deposition of particles with moderate $\mathrm{St}^+$ and large $\Gamma$. A potential approach, but beyond the scope of the current work, would be to ascribe some Stokes number dependence on the aerodynamic drag, which in principle makes sense, as these particles may be more efficient at extracting momentum from the background turbulent flow. 

In order to reconcile the phenomenological and phase-space approaches, the simplest modifications one could make to \eqref{de}, are to replace the diffusivity with a model for $\lambda$ and to add a correction to the Stokes settling velocity:
\begin{equation}
    -\lambda \frac{\partial C}{\partial z} - (v_g + \zeta)C = F, \label{corrected hoppel approach}
\end{equation}
where we have used and eddy-diffusivity-like closure, as well as a drift correction to the Stokes settling velocity, $\zeta = \gamma + \frac{\partial \lambda}{\partial z}$. This differential equation could be solved given adequate models for $\lambda$ and $\gamma$. A similar approach was taken by \citet{giardina_atmospheric_2019}, but the correction to the Stokes settling velocity was \textit{ad hoc}, and approached the limit of $v_g + u_\tau$ at large $\mathrm{St}^+$, which according to figure \ref{fig:mass_flux_and_sampling_dist}(b) is not true. Thus, in the case of negligible gradients of mean properties ($\frac{\partial C}{\partial z}$), such as for homogeneous isotropic turbulence, we should expect \eqref{corrected hoppel approach} to be able to reproduce the enhanced settling velocity seen in studies such as \cite{wang_settling_1993,rosa_settling_2016} due to the addition of the drift correction $\zeta$.

\section{Discussion and conclusions \label{discussion}}
In this work, we have coupled settling inertial Lagrangian point particles to direct numerical simulations of a turbulent boundary layer. Our goals were to identify how the independent variations in $\mathrm{Sv}^+$ and $\mathrm{St^+}$ modified the preferential sweeping mechanism and to use that insight to compare two fundamentally different approaches to modeling the settling characteristics of the inertial particles.  The first approach is based on the phase-space probability density of the ensemble of particles which allows us derive an exact formulation for the evolution of the average number concentration and the average particle momentum, while the second approach was a combination of a mass-conservation argument and a resistance based approach analogous to deposition models in current operational use. 

We found that variation in $\mathrm{Sv}^+$ modified the fluid velocities sampled by the particles, thus modifying the role that preferential sweeping plays in determining the settling velocity of the inertial particles. We hypothesize that particles with a given low or moderate $\mathrm{St}^+$ sample velocities closer to the cores of turbulent eddies leading to higher sampled downward velocities within the interior of the domain when $\mathrm{Sv}^+$ is higher, but not so high that gravity completely governs the trajectory of the particles. In that limit, they would simply move vertically through the turbulent eddies. When the particles attain a large enough $\mathrm{St}^+$, they become ballistic and uncorrelated with the flow velocities. At this point, the velocities sampled by the particles show a negligible variation due to changes in $\mathrm{Sv}^+$.

We showed that the fluid velocities sampled by the particles were approximately normally distributed in the log-layer. For the phase space approach,  this allowed us to represent the average fluid velocity sampled by the particles using a closure from \cite{reeks_continuum_1992} that involves a drift component proportional to the local concentration as well as a diffusion based closure proportional to the gradient of the concentration. By focusing on the dynamics within the logarithmic region of the turbulent boundary layer, we were able to use both the DNS data and an assumption that the turbulence is approximately locally homogeneous which allowed us to constrain the diffusive and drift components of the closure. We found that the drift component was nearly an order of magnitude larger than the diffusion closure for moderate $\mathrm{St}^+$ (when preferential sweeping is expected to be active), while both components were comparable in magnitude for the largest and smallest values of $\mathrm{St}^+$. Furthermore, the drift coefficient captured the variations in the average settling velocities due to variations in $\mathrm{Sv}^+$, thus capturing the modified preferential sweeping discussed in figure \ref{fig:Sv_sample_schematic}, whereas the diffusion component did not. We emphasize here that since the magnitude of the  drift component is nearly ten times that of the diffusion component for moderate Stokes numbers, the eddy-diffusivity assumption (which ignores the drift component \textit{a priori}) made in phenomenological models is fundamentally incomplete.  

To this end, the phase-space approach discussed above was compared to a phenomenological approach which utilized the eddy-diffusivity closure in order to estimate the turbulent diffusion of the inertial particles. In this model, the diffusion is parameterized in terms of the aerodynamic drag \citep{slinn_predictions_1980} and an empirical correction to the drag as a function of $\mathrm{St}^+$ \citep{zhang_size-segregated_2001}. We found that this approach represented the enhanced vertical settling velocity from the DNS in only a qualitative sense, underestimating the turbulent settling velocity by up to 500\% in some instances. We argued that this was due to the inappropriate scaling coefficients that the empirical approach relied on, as well as an inadequate inertial correction that failed to capture the moderate $\mathrm{St}^+$ behaviour, both of which were consequences of the inappropriate eddy-diffusivity closure for the vertical turbulent flux.

The reality is that models that attempt to represent particle deposition used in mesoscale and climate modeling are even simpler than the phenomenological model discussed above. For example, see \citet{kukkonen_review_2012} for a comprehensive review of deposition models in operational use for regional forecasting models in Europe. Some examples of the deposition models in current use are a simple application of Stokes law, slightly more complex resistance based models \citep{giardina_new_2018,emerson_revisiting_2020}, and simple gas-transfer deposition models based on \citet{slinn_predictions_1980}. So in some sense, the phenomenological model we have considered in this paper adheres to physics in a way that operational models do not, as none of the previous models attempt to conserve aerosol mass in any way and rely on empirical corrections to match existing data. 

However, these models are in common use because of their simplicity. While PDF approach discussed in \ref{pdf approach} is exact and extremely explicit in the way that it describes the role of each individual mechanism, it is complex in a way that is likely off-putting for many applied scientists. Thus, an attainable goal for the results in this work would be to use the complex phase space models to inform and approximate simpler resistance-based models for eventual operational use. We have attempted to do that by estimating $\lambda$ in \eqref{lambda} from the DNS. As there are plenty of existing models for $\langle w^2\rangle$ for a turbulent boundary layer (see \citet{smits_highreynolds_2011,kunkel_study_2006,marusic_similarity_1997} for example), and for $\tau$ (see \cite{sikovsky_particle_2019}), the turbulent diffusion component could be easily estimated. With an adequate model for $\gamma$ in a turbulent boundary layer, which itself is beyond the scope of the current work, a drift correction to the Stokes settling velocity could be easily implemented in a modified phenomenological model, and an example was given in \eqref{corrected hoppel approach}. This represents a simple but more accurate approximation that could be implemented in operational models.

         

\begin{acknowledgements}
The authors would like to acknowledge Grant No. W911NF2220222 from the U.S. Army Research Office. The authors would also like to thank the Center for Research Computing at the University of Notre Dame, as well Tim Berk. \newline The authors report no conflict of interest.
\end{acknowledgements} 

\bibliographystyle{spbasic_updated.bst} 
\bibliography{main_blm.bib}

\end{document}